\documentclass[journal]{IEEEtaes}
\usepackage{amsmath,amsfonts}
\usepackage[linesnumbered,ruled,vlined]{algorithm2e}
\usepackage{xcolor}
\usepackage[showdeletions]{color-edits}
\addauthor{JW}{magenta}
\usepackage{array}
\usepackage[caption=false,font=normalsize,labelfont=sf,textfont=sf]{subfig}
\usepackage{textcomp}
\usepackage{graphicx}
\usepackage{subfig}
\usepackage{mdframed}
\usepackage{xcolor}
\usepackage{cite}
\usepackage{color}
\usepackage{lineno}
\usepackage{bbm}
\usepackage{easyReview}
\usepackage{marginnote}
\usepackage{booktabs}
\usepackage{multirow}
\usepackage{mdframed}
\usepackage{xcolor}
\definecolor{red}{rgb}{0,0,0} 

\mdfdefinestyle{algstyle}{
    backgroundcolor=blue!10,
    linecolor=blue!200,
    roundcorner=2pt,
    innertopmargin=1pt
}
\usepackage[most]{tcolorbox} 
\newtcolorbox{subfigbox}[1][]{
    width=0.24\linewidth,        
    colback=yellow!200,          
    colframe=yellow!200, 
    boxrule=1pt,          
    arc=0pt,                
    boxsep=0pt,             
    left=0pt,               
    right=0pt,              
    top=0pt,                
    bottom=0pt,             
    before skip=0pt,        
    after skip=0pt,         
    enhanced,               
    #1 
}
\begin{document}
\title{Dimension Scaling SR-Net for Super-Resolution Radar Range Profiles}
      
\author{Ziwen~Wang}
\member{Graduate student Member,~IEEE}

\author{Jianping~Wang}
\member{Member,~IEEE}

\author{Pucheng~Li}
\member{Graduate student Member,~IEEE}

\author{Zegang~Ding}
\member{Senior Member,~IEEE}
\affil{Beijing Institute of Technology}

\receiveddate{Manuscript received XXXXX 00, 0000.\\
This work was supported in part by the National Science Foundation of China under Grant 62227901 and Grant 62101035, the Key Program of
the National Science Foundation of China under Grant 61931002. }
\corresp{ {\itshape (Corresponding author: Jianping Wang)}.}

\authoraddress{Ziwen Wang, Jianping Wang, Pucheng Li and Zegang Ding are with the School of Information and Electronics, Beijing Institute of Technology, and also with the Key Laboratory of Electronic and Information Technology in Satellite Navigation, Ministry of Education, Beijing 100081, China.
(e-mail: \href{mailto:3120235592@bit.edu.cn}{3120235592@bit.edu.cn}; \href{mailto:jianpingwang@bit.edu.cn}{jianpingwang@bit.edu.cn}; \href{mailto:pucklee1111@163.com}{pucklee1111@163.com}; \href{mailto:z.ding@bit.edu.cn}{z.ding@bit.edu.cn}).}

\maketitle

\begin{abstract}
  High-resolution radar range profile (RRP) is crucial for accurate target recognition and scene perception. To get a high-resolution RRP, many methods have been developed, such as multiple signal classification (MUSIC), orthogonal matching pursuit (OMP), and a few deep learning-based approaches. Although they break through the Rayleigh resolution limit determined by radar signal bandwidth, these methods either get limited super-resolution capability or work well just in high signal to noise ratio (SNR) scenarios. To overcome these limitations, in this paper, an interpretable unfolded neural network for super-resolution RRP (DSSR-Net) is proposed by integrating the advantages of both model-guided and data-driven models. Specifically, DSSR-Net is designed based on a sparse representation model with dimension scaling, and then trained on a training dataset. 
  Through dimension scaling, DSSR-Net lifts the radar signal into high-dimensional space to extract subtle features of closely spaced objects and suppress the noise of the high-dimensional features. It improves the super-resolving power of closely spaced objects and lowers the SNR requirement of radar signals compared to existing methods. The superiority of the proposed algorithm for super-resolution RRP reconstruction is verified via experiments with both synthetic and measured data.
\end{abstract}


\section{Introduction}

\IEEEPARstart{T}{hrough} emitting signals with a certain bandwidth, a radar system acquires the scattered echoes of the targets together with noise. The echoes with various time delays are then manipulated with signal processing approaches to obtain radar range profiles (RRP), which characterize the distribution and scattering properties of the targets along the distance \cite{xing2002properties,du2005radar,du2008radar,wang2019super}. The time delays of the echoes introduce phase shifts of the frequency spectra of the emitted signal. So, by analyzing the properties of the frequency spectra of the echoes, the RRP can be reconstructed.  
In practice, one class of widely used and effective signal processing approaches to create targets' RRP is fast Fourier transform (FFT)-based methods, which coherently accumulate the energy of different frequency spectra \cite{duhamel1990fast}. However, the RRP obtained with the FFT-based methods has limited resolution defined by the signal bandwidth and relatively high sidelobes induced by partially coherent accumulation of different frequency spectra \cite{skolnik1962introduction,kim2021high}.

To overcome the aforementioned limitations of the RRP obtained with traditional FFT-based methods, many super-resolution RRP methods have been proposed. They can be divided into three categories: spectral estimation methods, model-based reconstruction methods, and deep learning-based methods. Below, we will discuss each of them.

\textbf{1. Super-resolution spectral estimation methods}: In recent decades, many spectral estimation methods have been developed \cite{chen2010introduction}. Among them, multiple signal classification (MUSIC) \cite{zhang2010direction,waweru2014performance} and estimation of signal parameters using rotational invariance techniques (ESPRIT) \cite{roy1989esprit} are two of the most frequently used approaches, which estimate the signal spectral components by exploiting the non-correlation between the subspaces spanned by signal and noise components.  However, with the increase of the signal-to-noise ratio (SNR), the subspaces spanned by signal and noise components become gradually non-separable, which leads to a significant decrease in the accuracy of signal estimation \cite{xu1994beamspace}. Additionally, both algorithms require prior knowledge or an estimation of the number of targets for subspace decomposition. However, the number of targets may not be acquired in certain cases.
 
\textbf{2. Model-based reconstruction methods}: Model-based methods formulate the super-resolution RRP reconstruction as a parameter estimation problem by employing an observation model between radar echo and an RRP \cite{chiang2000model,zhu2021model}. The formulation of these methods generally leads to an ill-posed inverse problem for parameter estimation. To overcome the ill-posedness of the related estimation problems, some prior knowledge, for instance, sparsity \cite{potter2010sparsity}, total variation (TV) \cite{tang2020compressive}, low-rank \cite{qiu2019jointly}, is introduced as a constraint to limit the size of the solution space, which circumvents the requirement of the number of targets used in super-spectral estimation methods and leads to a sparsity-constrained optimization problem. Although these optimization problems are very difficult to be solved, some algorithms have been proposed to find near optimal solutions. Among them, greedy algorithms, such as orthogonal matching pursuit (OMP) \cite{tropp2007signal,pati1993orthogonal} are employed to select the variables with the highest correlation in the signal residual to form the target signal. However, this strategy is susceptible \textcolor{red}{to making commitments} to certain solutions too early, resulting in poor estimation accuracy. On the other hand, to make the problem of the sparsity-constrained RRP reconstruction easier to solve, it can be relaxed to be a convex optimization problem. Then the alternating direction method of multipliers (ADMM) \cite{boyd2011distributed,9007644} and the half-quadratic splitting (HQS) \cite{zoran2011learning} are suggested to iteratively solve the objective function. These algorithms are generally computationally efficient and converge to a near-optimal solution to the original optimization problem. However, they can only incorporate the prior knowledge that is easy to describe analytically, hindering the exploitation of some inherent prior features and limiting performance.

\textbf{3. Deep learning-based reconstruction methods}: Deep learning-based methods have demonstrated remarkable performance in various interdisciplinary tasks, leveraging hidden deep-seated features within data for precise classification and estimation \cite{pang2017convolution,zhang2013single,xin2020wavelet}.
In the direction of arrival (DOA) related spectrum estimation problem, an end-to-end network composed of several ResNet modules \cite{he2016deep} has been proposed to establish a mapping relationship from the input signal and spatial frequency spectrum through a training procedure with large amounts of data. 
On the other hand, DeepFreq introduced a linear and several convolutional modules for spectrum super-resolution, which could obtain better frequency extraction capabilities \cite{izacard2019data}. This method has been extended by introducing complex linear, ResNet modules, and an autoencoder block. Among them, CResFreq is the state-of-the-art for radar RRP super-resolution \cite{pan2021complex}.
Although the aforementioned end-to-end neural networks can learn effective mappings by learning the various features from a large volume of data, they are extremely data-hungry and work as a black-box, which makes them not applicable to safety-critical applications.
Another line of research explores unfolded networks, which unfold aforementioned model-based optimization algorithms (such as ADMM, HQS, or ISTA) into neural network architectures with trainable parameters \cite{wang2024single,klioui2025circulant,su2023real}. These networks retain the interpretability of traditional signal processing methods while gaining the adaptability of data-driven learning. However, these models typically learn along a single dimension of the training data, such as the range domain. This dimensional limitation restricts their ability to capture deeper cross-domain dependencies and complex feature interactions within the training data, thereby limiting their super-resolution capability for closely spaced objects.
In summary, it is essential to develop an interpretable neural network that can effectively learn across multiple dimensions of the training data for robust RRP estimation.

In this paper, an interpretable unfolded network with a dimension scaling architecture (named DSSR-Net) is proposed to achieve the super-resolution RRP based on a sparse representation model.
Firstly, radar echo is transformed into a high-dimensional space by a multi-linear operation module, thereby capturing subtle features of closely spaced objects. Then, multiple iterative stages, consisting of data consistency and feature extraction modules, are proposed to extract targets' features and suppress the noise in a high-dimensional space. Finally, the extracted high-dimensional features are weighted and combined to achieve the super-resolution of RRPs through a data dimension reduction module. Last but not least, \textcolor{red}{an efficient} strategy is proposed to ensure the effectiveness of network training by replacing the line spectra with their Gaussian blurred counterpart as labels of the related training data, and a theoretical explanation is provided. The contributions of this paper are as follows.
\begin{enumerate}
\item A sparse representation model with dimension scaling is proposed for the super-resolution of analytical RRP.
\item To implement this analytical model, an interpretable unfolded network named DSSR-Net is proposed to obtain super-resolution RRP. It is guaranteed to provide significant knowledge gains from interpretable sparse reconstruction model.
\item A Gaussian-blurred counterpart as a label is used to improve training efficiency, along with an intuitive explanation and analytical insight showing that it enlarges the gradient descent direction and enhances robustness.
\item A series of experiments are conducted to validate the efficacy of the proposed DSSR-Net.
\end{enumerate}

The rest of the paper is organized as follows: Section~\ref{section2} presents the radar signal model and related methods for estimating an RRP. In Section~\ref{section3}, an interpretable model for sparse representation with dimension scaling is proposed. On top of that, a network with dimension scaling for super-resolution RRP is devised, and its training strategy is then fully discussed.
In Section~\ref{section4}, the results of numerical simulations and real-data experiments are presented to illustrate the effectiveness of the proposed DSSR-Net. Section \ref{section5} concludes the paper and further plans.

\section{Signal model and related methods}
\label{section2}
In this section, an RRP signal model and a basic FFT-based processing are recommended, which provides a foundation for understanding the subject matter. Subsequently, a brief introduction of related methods for estimating an RRP is provided.  

\begin{figure}[b]
  \vspace{-0.6cm}
  \hspace{-0.5cm}\includegraphics[width =9.7cm, height =3.85cm]{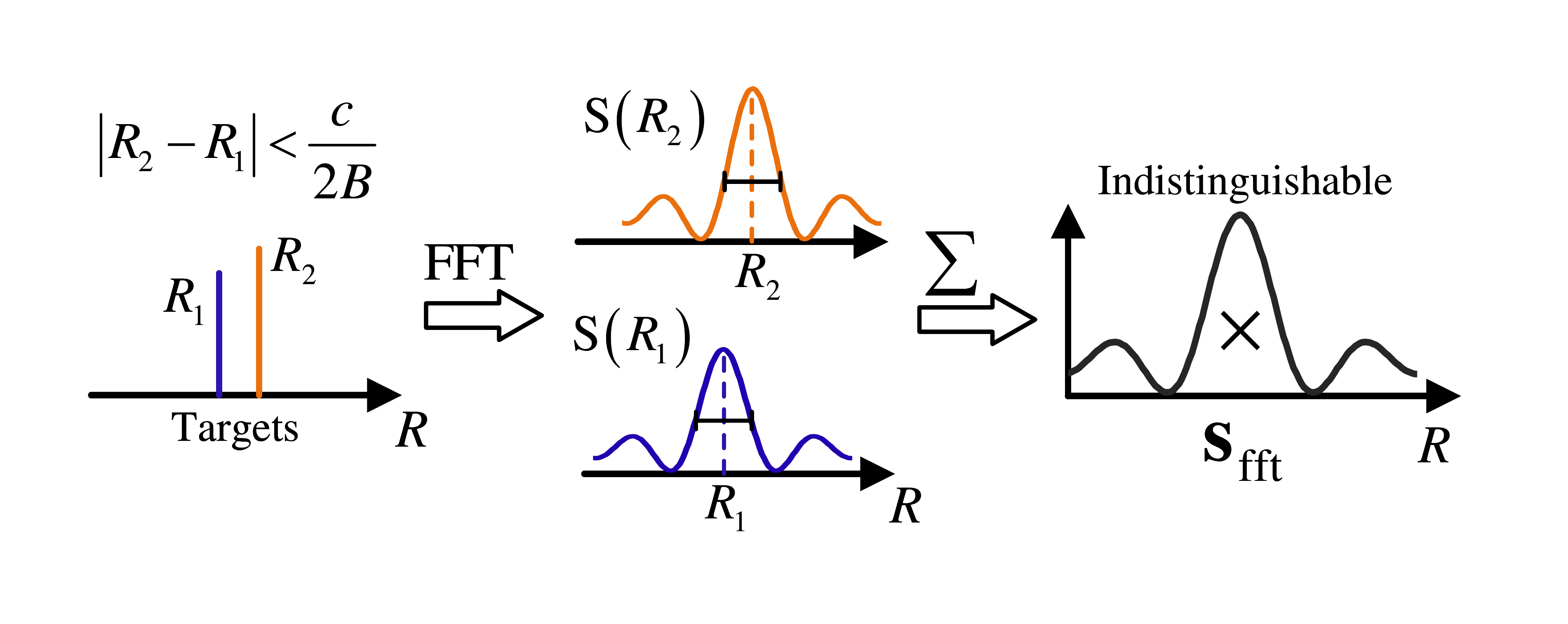}
  \vspace{-0.8cm}
  \caption{Illustration of the indistinguishable targets due to limited resolution of RRP obtained with FFT-based methods.}
  \label{fig:sinc}
\end{figure}
\subsection{Radar signal model}
A radar measures a target's range based on the delay of the received echo. The received echo can be expressed in the frequency domain as
\begin{align}
{\mathbf{y}} = \sum\limits_{p = 1}^P {{\sigma _p}H\left( {\boldsymbol{f}} \right)\exp \left( { - j2\pi {\boldsymbol{f}}{\tau _p}} { + {\varphi _p}}\right) + W\left( {\boldsymbol{f}} \right)} 
  \label{signal}
\end{align}
where ${\sigma _p}$, ${\tau _p} = 2{{{R_p}} \mathord{\left/
 {\vphantom {{{R_p}} c}} \right.
 \kern-\nulldelimiterspace} c}$, and ${{\varphi _p}}$ represent the scattering coefficient, the signal delay, and the initial phase of the $p^\text{th}$ target at range of  ${{R_p}}$, respectively. $c$ is the speed of light.
 $H\left(  \cdot  \right)$ is the frequency spectrum of the emitted signal and the rectangular window function ${\text{rect}}\left(  \cdot  \right)$ is commonly used. ${\boldsymbol{f}} \in \mathbb{R}{}^{N \times 1}$ denotes the vector of frequency sampling point, which can be discretely represented as ${\boldsymbol{f}} = {f_0} + {\boldsymbol{n}} \Delta f$, ${\boldsymbol{n}} = \left[ {0,1, \cdots ,N - 1} \right]$, ${f_0}$, $\Delta f$ and $N$ are the start frequency, frequency sampling interval, and the number of frequency sampling points, respectively. Naturally, the signal bandwidth is $B  = N\Delta f$ and $W\left( {\boldsymbol{f}} \right) \in \mathbb{C}{}^{N \times 1}$ is the additive white Gaussian noise (AWGN).

FFT-based methods extract the target range information by analyzing the frequency spectrum of the signal
\begin{equation}
{{\mathbf{x}}_{\text{f}}}{\text{ }} = {\text{FFT}}\left( {\mathbf{y}} \right){\text{ }} = A\sum\limits_{p = 1}^P {{\sigma _p}{F_H}\left[ {\pi \left( {{\mathbf{n}} - \frac{{2B}}{c}{R_p}} \right)} \right]}  + {\mathbf{w}}
\label{eq:fft}
\end{equation}
where $A$ is a complex constant; ${F_H}$ is the Fourier transform of $H\left(  \cdot  \right)$ and it is a $\mathrm{sinc}$ function when $H\left(  \cdot  \right) = {\text{rect}}\left(  \cdot  \right)$. $\mathbf{w} \in \mathbb{C}^{N \times 1}$ is the corresponding noise. One can see that the results of FFT-based methods are the superposition of a series of sinc function centered at $\frac{2B}{c}R_p$, where the peaks of main lobes of sinc functions indicate the positions of targets.
However, as the main lobe of a $\mathrm{sinc}$ function is with a certain width determined by the bandwidth $B$. It limits the range resolution of targets in the RRP obtained with FFT-based methods (illustrated in Fig.~\ref{fig:sinc}). To overcome the resolution limit caused by the signal bandwidth of FFT-based methods, super-resolution approaches are in high demand. The following will briefly introduce super-resolution methods based on models and deep learning.

\subsection{Data-driven deep learning-based reconstruction method}

\begin{figure}[t]
  \vspace{-0.8cm}
  \hspace{-0.3cm}
  \includegraphics[width =9.3cm, height =4.4cm]{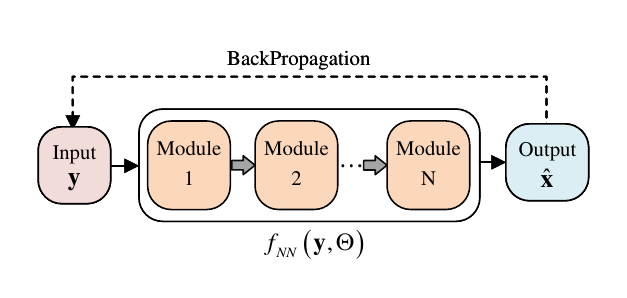}
  \vspace{-1cm}
  \caption{The simple schematic diagram of data-driven deep learning-based frequency estimation method.}
  \label{fig:DEEP}
\end{figure}

Data-driven deep learning-based reconstruction methods find the non-linear reconstruction mapping relationship between input and output by learning network parameters ${\boldsymbol{\Theta}}$ from massive training data. Specifically, as shown in Fig.~\ref{fig:DEEP}, the deep learning-based method usually inputs the radar echo ${\mathbf{y}}$ and outputs the estimation of ${{\mathbf{x}}}$ corresponding to each frequency. The network ${{f_{\text{NN}}}}$ usually consists of multiple modules, achieving an end-to-end frequency estimation. The objective function is
\begin{equation}
\mathop {\arg \min }\limits_\Theta  \sum\limits_{i = 1}^I {\mathcal{L}\left( {{f_\text{NN}}\left( {{{\mathbf{y}}_i},\Theta } \right),{{\mathbf{x}}_i}} \right)} 
\end{equation}
where $\mathcal{L}$ is the loss function, and mean square error (MSE) is generally used. $\left( {{\mathbf{y}_i},{\mathbf{x}_i}} \right),i=1,2,\cdots, I$ are the training data formed by radar echo and its associated RRP. Deep learning-based methods usually achieve good results. However, stacked network modules are usually not interpretable and can only be constructed based on experience. 

\subsection{Model-based Reconstruction method and its unfolded networks}

Besides the FFT-based methods, \eqref{signal} can be, by discretizing the targets' range ${{R_p}}$, represented as
\begin{equation} \label{eq:linearEq_discrete}
  {\mathbf{y}} = {\mathbf{Ax}} + {\mathbf{w}}
\end{equation}
where ${\mathbf{y}} \in {{\mathbb{C} }^{N \times 1}}$ represents the radar echo  and ${\mathbf{x}} \in {{\mathbb{C} }^{M \times 1}}$ is the target RRP, where $M$ denotes the number of range grids.
${\mathbf{A}} \in {{\mathbb{C} }^{N \times M}}$ is the observation matrix that represents the linear mapping relationship between the radar echo and the corresponding RRP, and its basic vector is $\exp \left( -j4\pi \boldsymbol{f} R_p/c  \right)$, and $\mathbf{w} \in \mathbb{C}^{N \times 1}$ is AWGN. 
As \eqref{eq:linearEq_discrete} is generally under-determined, it is generally solved by incorporating certain ``prior information", which is represented as  
\begin{equation}
{\mathbf{\hat x}} = \mathop {\arg \min }\limits_{\mathbf{x}} \frac{1}{2}\left\| {{\mathbf{y}} - {\mathbf{Ax}}} \right\|_2^2 + \lambda J\left( {{\mathbf{Dx}}} \right)
  \label{lasso}
\end{equation}
where $J\left( {{\mathbf{Dx}}} \right)$ denote the prior regularization term, which processes the transformation of $\mathbf{x}$ via domain transforming matrix $\mathbf{D}$ to a target domain, where the target prior information is utilized; $\lambda$ and $J\left(  \cdot  \right)$ are the penalty coefficient and the prior function of ${\mathbf{x}}$, respectively. 
In the scenario of signal sparsity, the prior regularization term commonly represents the widely employed ${L_1}$ norm, i.e., ${\left| {\mathbf{x}} \right|_1}$. The problem in \eqref{lasso} can be solved by convex optimization methods such as HQS. Taking HQS as an example, the specific solution goal is
\begin{equation}
\begin{gathered}
  \mathop {\arg \min }\limits_{\mathbf{x}} \frac{1}{2}\left\| {{\mathbf{y}} - {\mathbf{Ax}}} \right\|_2^2 + \frac{\rho }{2}\left\| {{\mathbf{x}} - {\mathbf{z}}} \right\|_2^2 \hfill \\
  \mathop {\arg \min }\limits_{\mathbf{z}} \lambda{\left| {\mathbf{z}} \right|_1} + \frac{\rho }{2}\left\| {{\mathbf{x}} - {\mathbf{z}}} \right\|_2^2 \hfill \\ 
\end{gathered} 
\label{hqs}
\end{equation}
where ${\mathbf{z}}$ is the introduced auxiliary variable. $\rho$ denotes the penalty parameter. The first step in \eqref{hqs} generally has a closed solution and alternately solves the two steps to obtain an iterative solution.
Specifically, the prior function corresponding to the sparse prior is the soft threshold function. However, choosing appropriate prior regularization terms and iterative hyperparameters is still a major difficulty in model-based reconstruction methods.

\begin{algorithm}[t]  
  \caption{Sparsity-Driven Unfolded Network Based on HQS.}
  \label{alg:alg1}
  \KwIn{$\mathbf{y}$, $K$, $k \gets 0$, ${{\mathbf{z}}^0} \gets \mathbf{0}$}
  \For{$k = 1$ \KwTo $K-1$}{
    ${{\mathbf{x}}^{k + 1}} = {\left( {{{\mathbf{A}}^H}{\mathbf{A}} + \rho ^ * {\mathbf{I}}} \right)^{ - 1}}\left( {{{\mathbf{A}}^H}{\mathbf{y}} + {{\rho ^ * }}{{\mathbf{z}}^k}} \right)$ \tcp*[r]{\color{blue}Data consistency iteration}
    ${{\mathbf{z}}^{k + 1}} = S\left( {{{\mathbf{x}}^{k + 1}},\frac{\lambda ^ * }{\rho ^ * }} \right)$ \tcp*[r]{\color{blue}Sparse prior iteration}
  }
  \Return $\mathbf{x}_K$
\end{algorithm}

Unfolding the above iterative process in \eqref{hqs} into an unfolded neural network offers a potential solution to acquire a set of suitable parameters through training. The processing procedure of the HQS-based sparsity-driven network is depicted in Algorithm \ref{alg:alg1}. However, the operation of unfolded networks along only a single dimension inherently restricts their capacity to capture complex, multi-dimensional feature interactions within the training data, thereby constraining their super-resolution performance for closely spaced targets.

To overcome the drawbacks of data-driven deep learning–based RRP reconstruction methods and the constraints of unfolded networks operating along a single dimension, a promising direction is to simultaneously leverage the interpretability of model-based approaches and the powerful capability of data-driven methods to capture multi-dimensional feature interactions. In the next section, we will present such an approach proposed to reconstruct super-resolution RRPs.
\section{DSSR-Net}
\label{section3}

In this section, a sparse representation model in high-dimensional space is presented with dimension scaling for the reconstruction of super-resolution RRP. Then, based on the presented model, an interpretable unfolded super-resolution network (named DSSR-Net) is devised, and the structure and function of each module are explained in detail. Finally, the training strategy of DSSR-Net is discussed.

\subsection{Sparse Representation Model with Dimension Scaling}
\label{section3.1}
In general, when two targets are in the same resolution unit, limited by the radar bandwidth, it will be difficult to distinguish them by using FFT-based methods and other methods that operate in single dimensional space.
To enhance the resolving power, a sparse representation model for super-resolution RRP is proposed. Specifically, the single-dimensional radar echo is first transformed into a high-dimensional RRP space to extract differential features in various dimensions. An iterative denoising model is then established in the high-dimensional space to extract the targets' features and suppress noise to obtain an accurate representation of RRP in high-dimensional space. Finally, the representation of RRP in high-dimensional space is projected to reconstruct the one-dimensional super-resolution RRP. The sparse representation model is described in detail below.

First, lifting the radar signal into high-dimensional space would be helpful to make subtle differences, i.e., frequencies, of targets in low-dimensional space easier to be separable and then be distinguished \cite{romero2021deepfilter}.
Therefore, in this paper, a multilinear operation is suggested to transform the measured one-dimensional signal into a high-dimensional RRP space, which can be expressed as
\begin{equation}\label{eq:linear}
{\mathbf{X}_f} = {\mathbf{F}}\left( {{{\boldsymbol{\theta }}_{\text{u}}}} \right)  {\mathbf{y}}
\end{equation}
where $\mathbf{F}\left(  \boldsymbol{\theta }_\text{u} \right) \in {\mathbb{C} ^{{C\times N \times N}}}$ represents the multilinear operation formed by stacking $C$ matrices with the dimensions of $N\times N$, and ${{\boldsymbol{\theta }}_\text{u}}$ represents the parameters of the operation. $\mathbf{X}_f\in \mathbb{C}^{C\times N}$ is the resultant high-dimensional signal after the operation. The process in \eqref{eq:linear} transforms the radar echo $\mathbf{y}$ into multiple RRPs and serves the role of an encoder, which learns multiple transformations similar to the Fourier transform in \eqref{eq:fft}. This potentially transforms the original signal into a high-dimensional RRP space, which allows for the focused extraction of different signal features and their fusion to achieve information integration \cite{ren2021adaptive}. 

However, a measured radar echo is practically contaminated by noise.
To alleviate the effect of noise and then reconstruct an accurate RRP, a denoising operation is necessary, which can be expressed as a consistency constraint in the high-dimensional space 
 \begin{equation}
 \mathop {\arg \min }\limits_{\mathbf{X}} \left\| {{\mathbf{X}_f} - {\mathbf{X}}} \right\|_F^2{\text{ }}
  \label{eq1}
 \end{equation}
where ${\left\|  \cdot  \right\|_F}$ denotes the Frobenius norm of a matrix. ${\mathbf{X}} \in {{\mathbb{C}}^{{C \times N}}}$ is the desired reconstructed RRP in the high-dimensional space.
To address the optimization problem in \eqref{eq1}, prior physical knowledge or features of the solution in high-dimensional space are generally exploited, which would lead to a proper and robust solution. By incorporating the prior term and the introduction of the auxiliary variable, \eqref{eq1} can be modified to
\begin{align}
\mathop {\arg \min }\limits_{\mathbf{X}} \left\| {{{\mathbf{X}}_f} - {\mathbf{X}}} \right\|_F^2 + \lambda {\mathcal{J}}\left( {\mathbf{Z}} \right)
  \quad s.t.\quad  {\mathbf{Z}}= \mathcal{D}\left( {\mathbf{X}} \right){\text{ }} 
  \label{zengguang}
\end{align}
where $\mathbf{Z}$ is introduced as an auxiliary variable of the prior information $\mathcal{D}(\mathbf{X})$. $\mathcal{D}\left( {\cdot} \right)$ represents the domain transformation function, which is similar to the linear matrix $\mathbf{D}$ in \eqref{lasso}, such as the discrete cosine transform (DCT) \cite{strang1999discrete} and the digital wavelet transform (DWT) \cite{zhang2019wavelet}, etc. 
$\mathcal{J}\left( \cdot  \right)$ is the prior penalty function, 
which generally involves non-linear processes, such as the soft threshold function, which can accomplish feature selection and noise suppression. 
The constraint term in \eqref{zengguang} is equivalently rewritten as ${\mathbf{X}} = \mathcal{T}\left( {\mathbf{Z}} \right)$, where $\mathcal{T}\left(  \cdot  \right)$ is an inverse function of $\mathcal{D}\left( {\cdot} \right)$. Note $\mathcal{T}\left(  \cdot  \right)$ can be considered as a domain transformation function.
Consequently, the unconstrained form of the objective function in \eqref{zengguang} is
\begin{equation}\label{eq:unconstrained}
\mathop {\arg \min }\limits_{{\mathbf{X}},{\mathbf{Z}}} \left\| {{{\mathbf{X}}_f} - {\mathbf{X}}} \right\|_F^2 + \lambda \mathcal{J}\left( {\mathbf{Z}} \right) + \mu \left\| {{\mathbf{X}} - \mathcal{T}\left( {\mathbf{Z}} \right)} \right\|_F^2
\end{equation}
where $\mu $ is the penalty coefficient. To solve the problem \eqref{eq:unconstrained}, it can be implemented by iteratively solving two optimization problems of $\mathbf{X}$ and $\mathbf{Z}$.
\begin{subequations}
\begin{align}
\hspace{-0.1cm}
    {{\mathbf{X}}_{k + 1}} &= \mathop {\arg \min }\limits_{\mathbf{X}} \left\| {{\mathbf{X}_f} - {\mathbf{X}}} \right\|_F^2 + \mu \left\| {{\mathbf{X}} - \mathcal{T}\left( {\mathbf{Z}_k} \right) } \right\|_F^2 \label{iterations1}
    \\[1mm]
    {{\mathbf{Z}}_{k + 1}} &=\mathop {\arg \min }\limits_{\mathbf{Z}} \lambda  \mathcal{J}\left( {\mathbf{Z}} \right) + \mu \left\| {{{\mathbf{X}}_{k + 1}} -  \mathcal{T}\left( {\mathbf{Z}} \right)} \right\|_F^2
  \label{iterations2}
\end{align}
\end{subequations}
where $k = 1,2, \ldots ,K$ is the index of an iteration. Concerning this two-step iteration process, a detailed discussion will be provided below.

\subsubsection{\textbf{First step}} 
The adjustment $\left\| {{\mathbf{X}} - \mathcal{T}\left( {\mathbf{Z}} \right)} \right\|_F^2$ in \eqref{iterations1} is uniformly penalized by the coefficient $\mu$. However, it is desirable to achieve a higher loss in the region of interest for the target. Therefore, it is helpful to employ an adaptive penalizing factor on each fitting deviation. This is somewhat similar to the attention mechanism in computer vision \cite{vaswani2017attention}, which puts the focus on regions of interest. Thus, \eqref{iterations1} can be revised as
\begin{equation}  \label{eq:X}
{{\mathbf{X}}_{k + 1}} = \mathop {\arg \min }\limits_{\mathbf{X}} \left\| {{\mathbf{X}_f} - {\mathbf{X}}} \right\|_F^2 + \left\| {{\mathbf{\Lambda }} \odot \left( {{\mathbf{X}} - \mathcal{T}\left( {\mathbf{Z}_k} \right)} \right)} \right\|_F^2
\end{equation}
where $\odot$ denotes the hadamard product, and ${\mathbf{\Lambda }} \in {\mathbb{R}^{C \times N}}$ is the adaptive weight matrix which replaces the constant $\mu$ in \eqref{iterations1}.

The solution of \eqref{eq:X} is obtained as follows. By setting its derivative to zero, one can get
\begin{align}\label{eq:X1}\nonumber
    {\mathbf{0}} &= \frac{{\partial \left[ {\left\| {{\mathbf{X}_f} - {\mathbf{X}}} \right\|_F^2 + \left\| {{\mathbf{\Lambda }} \odot \left( {{\mathbf{X}} -\mathcal{T}\left( {\mathbf{Z}_k} \right) } \right)} \right\|_F^2} \right]}}{{\partial {\mathbf{X}}}}
  \\[1mm]
   &= \left( {{\mathbf{X}} - {\mathbf{X}_f}} \right) + {{\mathbf{\Lambda }}^2} \odot \left( {{\mathbf{X}} - \mathcal{T}\left( {\mathbf{Z}_k} \right) } \right)
\end{align}
Rearranging the unknown ${\mathbf{X}}$ in \eqref{eq:X1} on one side leads to 
\begin{equation}
\left( {{\mathbf{1}} + {{\mathbf{\Lambda }}^2}} \right) \odot {\mathbf{X}} = {{\mathbf{X}}_f} + {{\mathbf{\Lambda }}^2} \odot \mathcal{T}\left( {\mathbf{Z}} \right)
\end{equation}
where $\mathbf{1} \in {\mathbb{R}^{C \times N}}$ is an all-ones matrix. The $l^\text{th}$ element in $\mathbf{X}$ is
\begin{align}
{\left[ {\mathbf{X}} \right]_l}  = \frac{{{{\left[ {\mathbf{X}_f} \right]}_l} + {{\mathbf{\Lambda }}_l^2}{{\left[ \mathcal{T}\left( {\mathbf{Z}_k} \right)\right]}_l}}}{{1 + {{\mathbf{\Lambda }}_l^2}}}= {\beta _l}{\left[ {\mathbf{X}_f} \right]_l} + \left( {1 - {\beta _l}} \right){\left[ { \mathcal{T}\left( {\mathbf{Z}_k} \right)} \right]_l}
\end{align}
where ${\left[  \cdot  \right]_l}$ denotes $l^\text{th}$ element of a matrix, and ${\beta _l} = 1/\left( {1 +{{\mathbf{\Lambda }}_l^2}} \right) \in \left( {0,1} \right)$. Then, the closed solution of \eqref{eq:X} is given by
\begin{equation}  \label{eq:x_iteration}
  {{\mathbf{X}}_{k + 1}} = {\boldsymbol{\beta }} \odot {\mathbf{X}_f} + \left( {{\mathbf{1}} - {\boldsymbol{\beta }}} \right) \odot  \mathcal{T}\left( {\mathbf{Z}_k} \right) 
\end{equation}
where ${\boldsymbol{\beta }} \in {\mathbb{R}^{C \times N}} $ is the weight matrix. As the adaptive penalizing matrix $\mathbf{\Lambda}$ is related to ${\mathbf{X}_f}$ and ${\mathcal{T}\left( {{{\mathbf{Z}}_k}} \right)}$, so does ${\boldsymbol{\beta }}$.
Therefore, an operator $\mathcal{B}\left(  \cdot  \right)$ with inputs being both of them is proposed to generate the weight matrix ${\boldsymbol{\beta }}$.

\subsubsection{\textbf{Second step}}

The prior regularization iterations in \eqref{iterations2} can be viewed as a process for extracting reasonable features, which can be represented using a proximal algorithm. This process takes multidimensional signals, denoted as $ {{\mathbf{X}}_{k + 1}}$, as input and produces the output of the proximal feature extraction
\begin{equation}
  {{\mathbf{Z}}_{k + 1}} =\mathcal{E}\left( {{{\mathbf{X}}_{k + 1}};{{\boldsymbol{\theta }}_\text{d}}} \right)
  \label{eq:Z1}
\end{equation}
where ${\mathcal{E}}\left(  \cdot  \right)$ denotes the proximal feature extraction operation, and ${{\boldsymbol{\theta }}_\text{d}}$ is the corresponding parameters.

By iteratively taking the operations in \eqref{eq:x_iteration} and \eqref{eq:Z1}, the representation in the high-dimensional RRP space $\mathbf{X}$ is obtained. Then, it is projected to the one-dimensional RRP to obtain the super-resolution RRP, given by 
\begin{equation}\label{final}
    {{\mathbf{x}}_{K}} = {{\mathbf{F}}^{ - 1}}\left( {{{\mathbf{X}}_{K}};{{\boldsymbol{\theta }}_\text{l}}} \right)
\end{equation}
where ${{\mathbf{F}}^{ - 1}}$ represents the dimensionality reduction operation, which represents an opposite operation to ${\mathbf{F}}$ in terms of data dimensions, and ${{{\boldsymbol{\theta }}_\text{l}}}$ is its parameter. Note that ${\left(  \cdot  \right)^{ - 1}}$ does not denote its inverse.

\begin{figure}[t]
\centering
  \vspace{-0.5cm}
  \includegraphics[width =5cm, height =10cm]{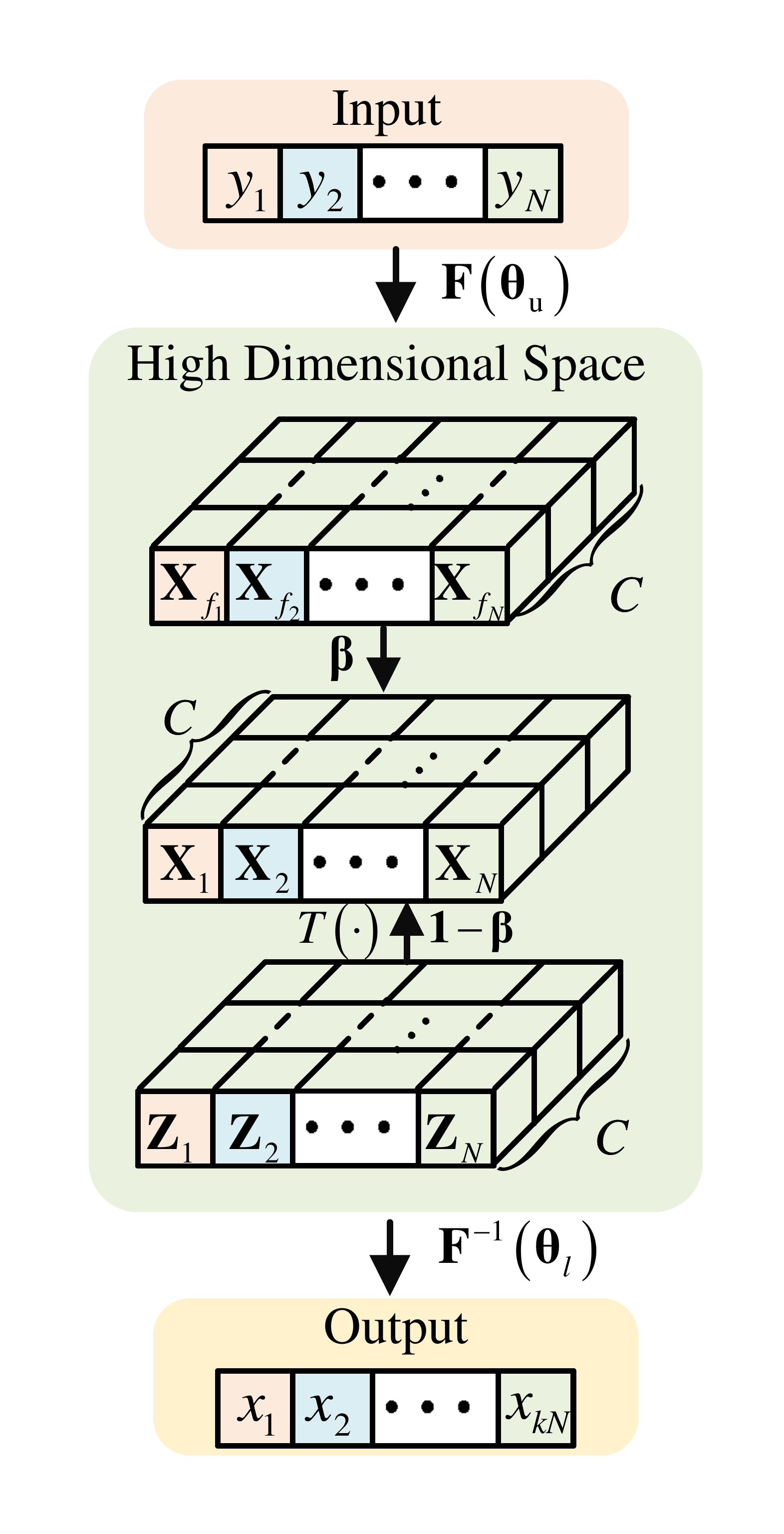}
  \vspace{-0.5cm}
  \caption{The diagram depicting the mapping relationships of variables in the proposed sparse representation model with dimension scaling.}
  \label{fig:mapping}
\end{figure}

The sparse representation model with dimension scaling for super-resolution RRP is outlined in Algorithm \ref{alg:alg2}. And the mapping relationships of all variables in the sparse representation model are illustrated in Fig.~\ref{fig:mapping}. It establishes a comprehensive theoretical framework for super-resolution RRP.
To implement this theoretical framework, a neural network architecture can be adopted, which will be presented below.

\begin{algorithm}[t]
  \caption{The sparse representation model with dimension scaling for super-resolution RRP.}
  \label{alg:alg2}
  \KwIn{$\mathbf{y}$, $K$, $k \gets 0$, ${{\mathbf{Z}}_0} \gets \mathbf{0}$}
  $\mathbf{X}_f \gets \mathbf{F} (\boldsymbol{\theta }_\text{u}) \mathbf{y}$ \tcp*[r]{\color{blue}Echo dimension upsampling}
  \For{$k = 1$ \KwTo $K$}{
    $\boldsymbol{\beta } \gets \mathcal{B}(\mathbf{X}_f, \mathcal{T}(\mathbf{Z}_k))$ \tcp*[r]{\color{blue}Weight matrix generation}
    $\mathbf{X}_{k + 1} \gets \boldsymbol{\beta } \odot \mathbf{X}_f + (\mathbf{1} - \boldsymbol{\beta }) \odot  \mathcal{T}(\mathbf{Z}_k)$ \tcp*[r]{\color{blue}Data consistency}
    ${{\mathbf{Z}}_{k + 1}} \gets \mathcal{E}(\mathbf{X}_{k + 1};\,{\boldsymbol{\theta }}_\text{d})$ \tcp*[r]{\color{blue}Proximal feature extraction}
  }
  $\boldsymbol{x}_K \gets \mathbf{F}^{ - 1}( \mathbf{X}_K; \boldsymbol{\theta}_\text{l})$ \tcp*[r]{\color{blue}Dimension reduction}
  \Return $\mathbf{x}_K$
\end{algorithm}

\begin{figure*}[t]
    \centering
  \includegraphics[width =16.8cm, height =11cm]{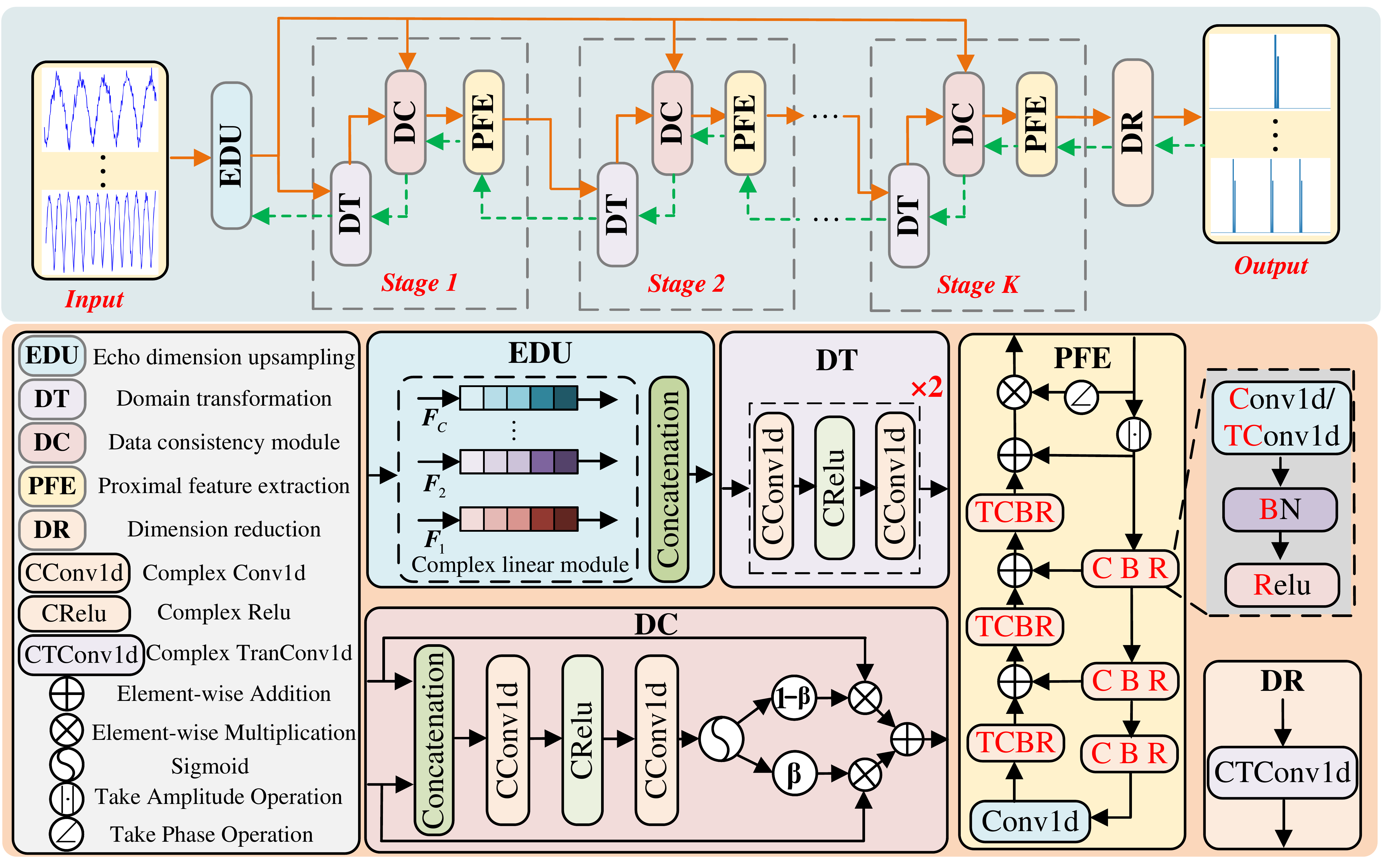}
  \caption{The schematic diagram of the proposed DSSR-Net. At the beginning and end, the EDU and the DR represent the echo dimension upsampling and the dimension reduction module, respectively. 
  Within these $K$ sub-iteration stages, DT, DC, and PFE are the domain transformation module, data consistency module, and proximal denoising module, respectively.}
  \label{fig:proposed}
\end{figure*}

\subsection{DSSR-Net: A Convolutional Neural Network with Dimension Scaling for Super-Resolution RRP} \label{DSSR-Net}
A convolutional Neural Network designed with dimension scaling for super-resolution RRP is presented in this section. Its overall flowchart is illustrated in Fig.~\ref{fig:proposed}. The network consists of an Echo Dimension Upsampling (EDU) module, $K$ sub-iteration stages, and a Dimension Reduction (DR) module. Each sub-iteration stage includes three distinct modules: the Domain Transformation (DT) module, the Data Consistency (DC) module, and the Proximal Feature Extraction (PFE) module. The sub-iteration stage is the key part of the unrolled network structure, enabling interpretable iterative processing within the high-dimensional feature space.
The specific structure of these modules will be provided in detail in the following.
\subsubsection{\textbf{Echo dimension upsampling (EDU) module}}
The schematic diagram of the EDU module is illustrated in the `EDU' block of Fig.~\ref{fig:proposed}. The input echoes are processed through a complex linear module with $C$ channels to implement the multilinear operation in \eqref{eq:linear}. Through the $i^\text{th}$ channel, a linear transformation is performed by ${{\mathbf{F}}_i}\mathbf{y}$, where ${{\mathbf{F}}_i}\in {{\mathbb{C}}^{{N \times N}}}$.
By synthesizing all the outputs of $C$ channels, the high-dimensional representation $\mathbf{X}_f$ can be obtained as
\begin{equation}
  {\mathbf{X}_f} = \mathrm{Cat}\left[ {{{\mathbf{F}}_1}{\boldsymbol{y}},{{\mathbf{F}}_2}{\boldsymbol{y}}, \ldots ,{{\mathbf{F}}_C}{\boldsymbol{y}}} \right]
\end{equation}
where $\mathrm{Cat}\left[  \cdot  \right]$ represents the channel concatenation operation.
This process enables the identification of different features of closely spaced targets in high-dimensional space, thereby facilitating super-resolution.

\subsubsection{\textbf{Domain transformation (DT) module}}
For super-resolution RRP reconstruction, the transformation $\mathcal{T}\left( {\cdot} \right)$ could be inverse of DCT, DWT, which is determined based on $\mathcal{D}\left( {\cdot} \right)$. Usually, it can only be empirically selected, which may not get satisfactory results. Therefore, it is necessary to design a learnable $\mathcal{T}\left( {\cdot} \right)$ to mine transform domain features from data. Here, we use two complex 1D convolution layers with a complex ReLU layer in between to build this module, resulting in the output feature variable $\mathcal{T}\left( {{{\mathbf{Z}}_k}} \right)$. This resultant module is shown as the `DT' block in Fig.~\ref{fig:proposed}.

\subsubsection{\textbf{Data consistency (DC) module}}
\label{section3.2.3}
The analysis presented in \textbf{First step} of Sec~\ref{section3}-\ref{section3.1} indicates that $\boldsymbol{\beta}$ depends on  ${\mathbf{X}_f}$ and ${\mathcal{T}\left( {{{\mathbf{Z}}_k}} \right)}$. 
To make it adaptive to various datasets, we construct a neural network module to learn $\boldsymbol{\beta}$ from ${\mathbf{X}_f}$ and ${\mathcal{T}\left( {{{\mathbf{Z}}_k}} \right)}$. Specifically, the output ${\mathbf{X}_f}$ of the EUD module and the output ${\mathcal{T}\left( {{{\mathbf{Z}}_k}} \right)}$ of the DT module are concatenated firstly, and followed by a combination of a complex convolution layer, a complex ReLU layer, another complex convolution layer and a sigmoid layer to generate $\boldsymbol{\beta}$.
Then, $\boldsymbol{\beta}$ and $\mathbf{1}-\boldsymbol{\beta}$ are inserted into \eqref{eq:x_iteration} to get ${{\mathbf{X}}_{k + 1}}$. This module is shown as the `DC' block in Fig.~\ref{fig:proposed}. 


\subsubsection{\textbf{Proximal feature extraction (PFE) module}}
For super-resolution RRP reconstruction,
the proximal operator ${\mathcal{E}}\left(  \cdot  \right)$ in \eqref{eq:Z1} extracts subtle differences of features of closely spaced targets. In practice, the amplitudes of the RRP indicates if targets exist at different distances, which usually has a certain prior distribution. By contrast, the phases of RRP are randomly distributed and \textcolor{red}{have no special} prior information. Motivated by this fact, we use a learnable operator ${\mathcal{\Tilde{E}}}$ to extract the amplitude features of the high-dimensional representation ${{\mathbf{X}}_{k + 1}}$ of the RRP, and preserve its corresponding phases, which can be expressed as
\begin{equation}
{{\mathbf{Z}}_{k + 1}} = {\mathcal{\Tilde{E}}}\left( {\left| {{{\mathbf{X}}_{k + 1}}} \right|;{{\boldsymbol{\theta }}_{\text{d}}}} \right) \odot {\text{angle}}\left( {{{\mathbf{X}}_{k + 1}}} \right)
  \label{eq:Z}
\end{equation}
where $\left|  \cdot  \right|$ and $\mathrm{angle}\left(  \cdot  \right)$  represent the modulus operation and the phase extraction, respectively. The proximal operation in \eqref{eq:Z} performs a filtering-like process. It can be implemented by the `Unet' structure, which has achieved significant success in image processing \cite{ronneberger2015u,siddique2021u}. So, in this paper, \textcolor{red}{we also adopt} the `U-net' structure to implement the PFE module. Specifically, the PFE module is formed by several `CBR' and `TCBR' blocks, each including a 1D convolution layer or 1D transposed convolution layer, a batch normalization layer, and a ReLU activation layer as illustrated in the `PFE' block of Fig.~\ref{fig:proposed}.  

\subsubsection{\textbf{Dimension reduction (DR) module}}

Following several iterations, the output ${{\mathbf{X}}_{k + 1}}$ is a high-dimensional representation of RRP. A module ${{\mathbf{F}}^{ - 1}}$ has to be constructed to combine the extracted high-dimensional representation in a weighted manner to obtain super-resolution RRP (as shown in \eqref{final}). 
In this paper, this module is implemented using a simple complex 1D transposed convolution layer.


\subsection{Network Training}
\label{sec_3_3_1}

To train this network, a training dataset comprising 500,000 sets of computer synthetic data is generated according to \eqref{signal}. The parameters used to construct the training dataset are listed in Table~\ref{table:Distribution}, where $ \mathcal{U} \left( a,b\right)$ represents a uniform distribution within the interval $\left( a,b\right)$.

Furthermore, the target ranges are set according to a Gaussian distribution, i.e., after randomly selecting the range of the first target, the subsequent targets have a spacing that follows a Gaussian distribution with zero mean and a variance of ${3 \mathord{\left/
 {\vphantom {3 N}} \right.\kern-\nulldelimiterspace} N}$. Considering the trade-off between operational efficiency and network signal reconstruction capability, the number of sub-iteration stages in the network is chosen to be three through experimental tuning.
The network was trained for 100 epochs using the Adam optimization algorithm with a learning rate of $5 \times 10^{-4}$ and a batch size of 1024.

\begin{table}[t]
  \caption{The training dataset parameters}   
      \centering
      \hspace{-0.5cm}
  \begin{tabular}{ccc}
  \toprule
    \textbf{Parameter} &  \textbf{Distribution} &\textbf{Unit}	 \\
    \midrule
    Range points & $N$ = 64& -    \\
    Component number &  ${P} \sim \mathcal{U} \left( {1,10 } \right)$ & - \\
    Initial phase &  ${\varphi _p} \sim \mathcal{U} \left( {0,2\pi } \right)$ & rad\\
    Range &   $R_p \in \left( {0,{c \mathord{\left/
 {\vphantom {c {2\Delta f}}} \right.
 \kern-\nulldelimiterspace} {2\Delta f}}} \right)$ & m\\
    Amplitude &   ${\sigma _p} \sim \mathcal{U} \left( {0.1,1} \right)$ & - \\
    Signal SNR & $\text{SNR}\sim \mathcal{U} \left( {0,50} \right)$ & $\mathrm{dB}$\\
  \bottomrule
  \end{tabular}
  \label{table:Distribution}
\end{table}

\begin{figure}[t]
  \centering
  \vspace{-0.4cm}
    \includegraphics[width =8cm, height =6.1cm]{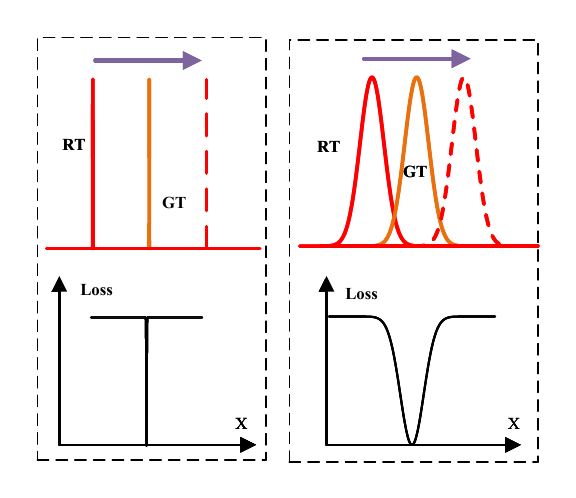}
   \vspace{-0.5cm}
  \caption{The impact of training loss for line spectra and the corresponding Gaussian blurred counterpart.}
  \label{fig:gaussian}
\end{figure}

The overall loss function for the entire network training is 
\begin{equation}
\mathcal{L} = \sum\nolimits_{p = 1}^P {\left\| {{\mathbf{R}}{{\mathbf{T}}_p} - {\mathbf{G}}{{\mathbf{T}}_p}} \right\|_2^2}  
\end{equation}
where ${{\mathbf{G}}{{\mathbf{T}}_p}}$ and ${{\mathbf{R}}{{\mathbf{T}}_p}}$ represent the ground truth (GT) of the line spectra and the network's estimated result of the $p^\text{th}$ data, respectively.
It is important to note that the GT used for training in this paper is not the traditional line spectra, which represent the exact positions of targets.
Instead, a Gaussian blurred counterpart of line spectra is employed, which was first used in \cite{izacard2019data}, and the similar idea has also appeared in other fields~\cite{1003065,1176936}. In this work, we provide an intuitive explanation and analytical insight: the blurred kernel enlarges the gradient descent direction and enhances the robustness of training. In addition, we found that, aside from the Gaussian blurred kernel, alternative blurred kernels such as Laplace, Lorentzian, and Triangular can also achieve similar effects. More details will be provided in the supplementary material. 
As illustrated in the left panel of Fig.~\ref{fig:gaussian}, if a line spectrum is used as a GT, gradients of the loss might be diminishing in the majority of intervals, which will make it very difficult to effectively update the network parameters. 
By contrast, utilizing a Gaussian blurred counterpart of line spectra significantly alleviates this issue, guiding the neural network to descend along the loss direction in theory, thus accelerating and aiding convergence (see the right panel of Fig.~\ref{fig:gaussian}). \textcolor{red}{Note that, since the imaging grid is discrete, when the true frequencies do not align with the sampled points, the amplitudes of the Gaussian blurred labels may become smaller than the unit amplitude. This could adversely affect the network’s training. To address this issue, we normalize the sampled Gaussian blurred labels such that the maximum amplitude of each Gaussian kernel is set to the unit one. Meanwhile, we want to mention that this strategy cannot completely eliminate the off-grid error. To complete overcome this issue, further investigations will be taken in future.}

\begin{figure}[t]
  \centering
  \vspace{-0.4cm}
  \includegraphics[width =8.4cm, height =3.8cm]{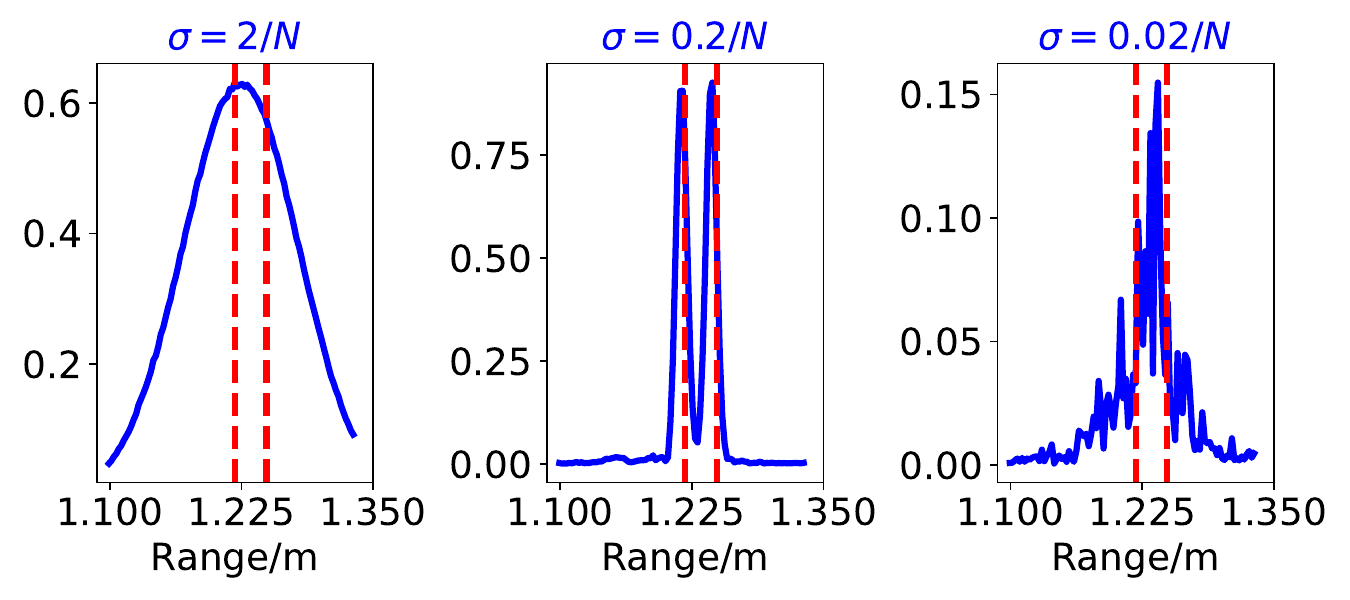}
  \vspace{-0.5cm}
  \caption{Output results using different widened Gaussian kernel variances of $\sigma = {2 \mathord{\left/
 {\vphantom {2 N}} \right.\kern-\nulldelimiterspace} N}$, $\sigma = {0.2 \mathord{\left/
 {\vphantom {2 N}} \right.\kern-\nulldelimiterspace} N}$, and $\sigma = {0.02 \mathord{\left/
 {\vphantom {2 N}} \right.\kern-\nulldelimiterspace} N}$.}
 \vspace{-0.5cm}
  \label{fig:gs}
\end{figure}
 
To experimentally verify the effect of the variance of the Gaussian kernel on the resolution performance, two targets are placed within a Rayleigh resolution unit.
Fig.~\ref{fig:gs} illustrates the network output results for three Gaussian kernels with different variances, namely ${2 \mathord{\left/
 {\vphantom {2 N}} \right.\kern-\nulldelimiterspace} N}$, ${0.2 \mathord{\left/
 {\vphantom {2 N}} \right.\kern-\nulldelimiterspace} N}$, and ${0.02 \mathord{\left/
 {\vphantom {2 N}} \right.\kern-\nulldelimiterspace} N}$, respectively, 
where ${1 \mathord{\left/{\vphantom {2 N}} \right.\kern-\nulldelimiterspace} N}$ represents a Rayleigh resolution unit. A variance of ${0.02 \mathord{\left/{\vphantom {2 N}} \right.\kern-\nulldelimiterspace} N}$ corresponds to about ${1 \mathord{\left/
 {\vphantom {1 {50}}} \right.\kern-\nulldelimiterspace} {50}}$ resolution unit, which exceeds the practical super-resolution capacity. Such an excessively narrow kernel is a more accurate approximation of a $\delta $ function, but overly restricts the gradient direction, adversely affecting the training process.
 Nevertheless, using a Gaussian kernel with an excessively large variance will inevitably result in a degradation of the resolution, thereby rendering it impossible to distinguish the closely-spaced targets. Therefore, a Gaussian kernel with a proper variance has to be found. In this paper, the variance of the Gaussian kernel is set ${0.2 \mathord{\left/
 {\vphantom {2 N}} \right.\kern-\nulldelimiterspace} N}$ to obtain a good result. It supports a moderate level of super-resolution enhancement while still maintaining effective gradient propagation during training.
 \textcolor{red}{In addition, under this setting, the aforementioned off-grid error caused by the misalignment between the true frequency and the sampled frequency point is already very small and has only a marginal effect on network training.}

\section{Experiments}
\label{section4}
In this section, numerical simulations and experimental measurements were conducted to analyze the performance of the proposed super-resolution RRP reconstruction method. 
To validate the superiority of the proposed DSSR-Net, comparisons were made with established methods, i.e., FFT \cite{duhamel1990fast}, MUSIC \cite{zhang2010direction}, OMP \cite{pati1993orthogonal}, ADMM with 100 iterations \cite{boyd2011distributed}, ADMM-Net \cite{wang2024single}, DeepFreq~\cite{izacard2019data}, and CResFreq, which is the state-of-the-art method in RRP super-resolution \cite{pan2021complex}. For fair comparison, all neural network-based baseline methods were trained for 100 epochs and used the same dataset as ours. Specifically, ADMM-Net was implemented with 60 layers, a learning rate of 0.001, and optimized using the Adam optimizer. Both DeepFreq and CResFreq were trained with the same settings as reported in their original papers, and the model parameters of DSSR-Net are shown in Table \ref{table:Model}. In this setting, $K$ denotes the kernel size, ${\text{Stride}}$ represents the convolution step size, and ${C_{in}}$ and ${C_{out}}$ correspond to the number of input and output feature channels, respectively.

\subsection{Super-resolution RRP performance analysis}
\label{section4.1}
The most direct measure of super-resolution RRP performance is the ability to resolve two closely spaced targets, ensuring not only their separation but also maintaining some precision in their relative and absolute positions. In this paper, the successful super-resolution RRP is defined as follows
\begin{equation}
  \left\{ {{x_{{p_1}}},{x_{{p_2}}}} \right\} \in {\Omega _1} \cap {\Omega _2} \cap {\Omega _3} \cap {\Omega _4}
\end{equation}
where ${x_{{p_1}}},{x_{{p_2}}}$ represent two resolved points, and only two points are allowed.
\begin{align}\nonumber
    {\Omega _1} &= \left\{ {{x_{{p_i}}}\left| {{x_{{p_i} - 1}} < {x_{{p_i}}} > {x_{{p_i} + 1}}} \right.} \right\},
    \\[1mm]
    \nonumber
   {\Omega _2} &= \left\{ {{x_{{p_i}}}\left| {{x_{{p_i}}} > {{\max \left( {\mathbf{x}} \right)} \mathord{\left/
 {\vphantom {{\max \left( {\mathbf{x}} \right)} 2}} \right.
 \kern-\nulldelimiterspace} 10}} \right.} \right\},\\[1mm]
    \nonumber
    {\Omega _3} &= \left\{ {{x_{{p_i}}}\left| {\Delta x - \frac{\rho }{4} < \left| {{\text{idx}}\left( {{x_{{p_1}}}} \right) - {\text{idx}}\left( {{x_{{p_2}}}} \right)} \right| < \Delta x + \frac{\rho }{4}} \right.} \right\},\\[1mm]
    {\Omega _4}& = \left\{ {{x_{{p_i}}}\left| {\left| {{\text{idx}}\left( {{x_{{p_i}}}} \right) - {\text{idx}}\left( {{x_{{p_{i\_t}}}}} \right)} \right| < \frac{\rho }{4}} \right.} \right\}.
\end{align}
 They denote the set of conditions for successful super-resolution RRP, and the specific explanation is as follows
\begin{enumerate}
\item ${\Omega _1}$ represents the condition that the extracted point is a peak, i.e., it is greater than the points to its left and right. 
\item ${\Omega _2}$ represents the condition that the extracted point is greater than 0.1 peak value. In this paper, after normalization, this value is typically 0.1. 
\item ${\Omega _3}$ indicates that the distance between the two extracted points is between $\Delta {x_n} - \frac{\rho }{4}$ and $\Delta {x} + \frac{\rho }{4}$, where $\Delta {x}$ represents the true distance between the two points. $\mathrm{idx}\left( \cdot \right)$ represents the range corresponding to the extracted points, ensuring the relative distance accuracy of the two points after super-resolution.
\item ${\Omega _4}$ ensures the absolute positional accuracy of the points after resolution, with $\mathrm{idx}\left( {{x_{{p_i}}}} \right)$ representing the range of the extracted point and $\mathrm{idx}\left( {{x_{pi\_t}}} \right)$ its true range, ensuring that it is less than $\frac{\rho }{4}$.
\end{enumerate}

\begin{table}[t]
  \caption{Network parameters of DSSR-Net} 
  \centering
  \begin{tabular}{cc}
  \toprule
    \textbf{Modules} &  \textbf{Parameters}	 \\
    \midrule
    EDU & $C=64$    \\[0.5mm]
    DT   &  $K=3,\text{Stride}=1, C_{in}=64, C_{out}=64$  \\[0.5mm]
    DC  &  $K=3,\text{Stride}=1, C_{in}=64, C_{out}=64$  \\[0.5mm]
    \multirow{2}{*}{PFE}  &  $K_{in}=3, \text{Stride}_{in}=2, C_{in}=64,128,256$\\
         & $K_{out}=3,\text{Stride}_{out}=2,C_{out}=256,128,64$\\[0.5mm]
    DR  &   $K=25,Stride=16,C_{in}=64,C_{out}=1$\\
  \bottomrule
  \end{tabular}
  \label{table:Model}
\end{table}

The phase transition analysis based on two points is performed to intuitively measure the robustness of the algorithm \cite{donoho2006breakdown,donoho2010precise,10057426}. Specifically, based on the above definition of successful super-resolution, we statistically analyze the probability of successful super-resolution for different methods under varying SNR and different separations. A higher probability indicates a higher super-resolution capability. Here we define a super-resolution index coefficient, the relative distance ${\rho _\text{d}}$, which is the ratio of the target distance to the Rayleigh resolution ${\rho _\text{d}} = \frac{{\Delta {x_t}}}{\rho }$. In our experiments,  
the SNR of radar echo varies between 0 and 26 $\,\mathrm{dB}$, while ${\rho _\text{d}}$ is set between 0.5 and 1.1. 
Each experiment is conducted 1000 times using Monte Carlo simulations to ensure that the results obtained are statistically significant.

\begin{figure*}[t]
  \vspace{-0.5cm}
  \hspace{0.05cm}
  \subfloat[]{\includegraphics[width =4.1cm, height =3.6cm]{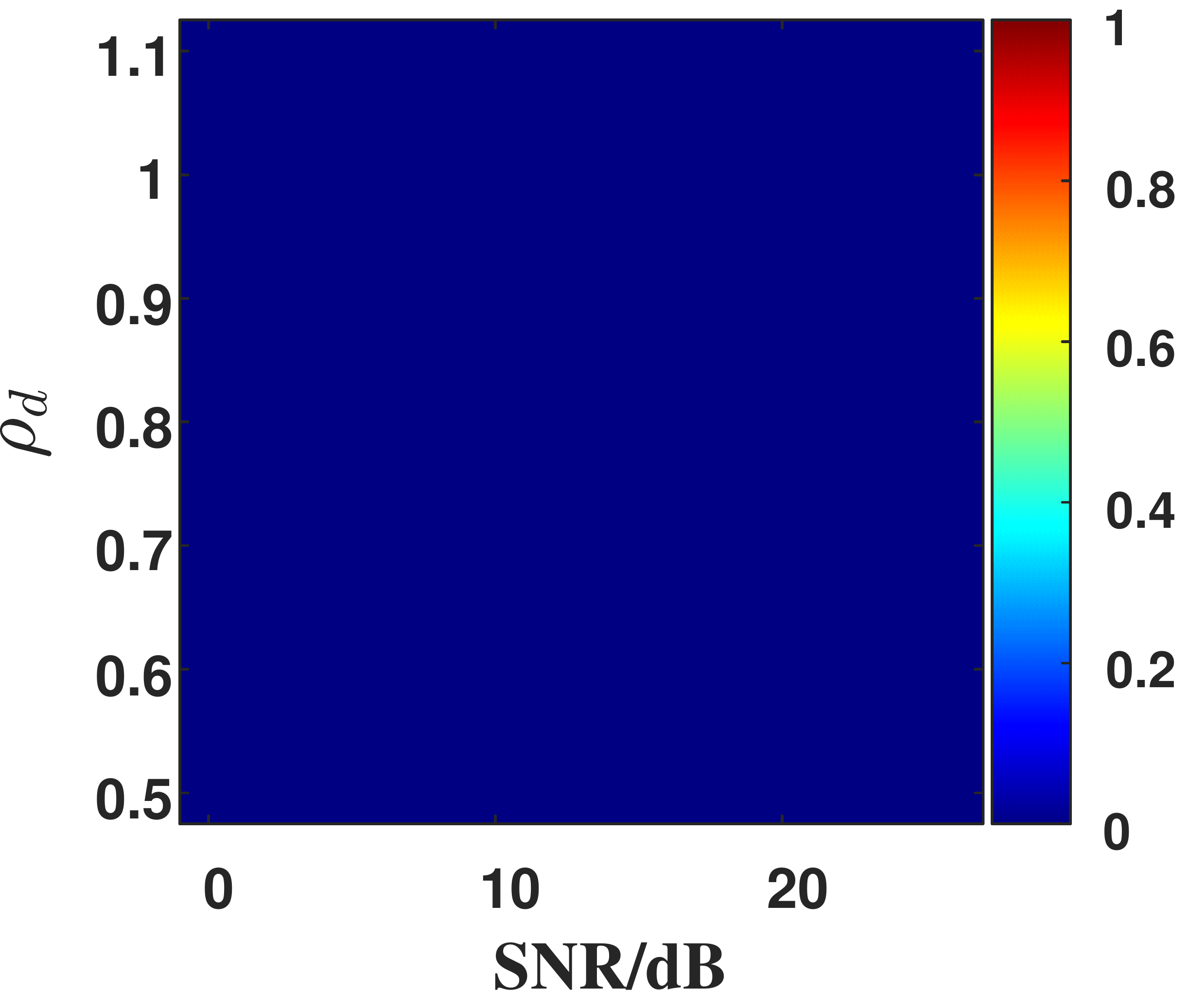}\label{FFT}}
  \hspace{0.05cm}
  \subfloat[]{\includegraphics[width =4.1cm, height =3.6cm]{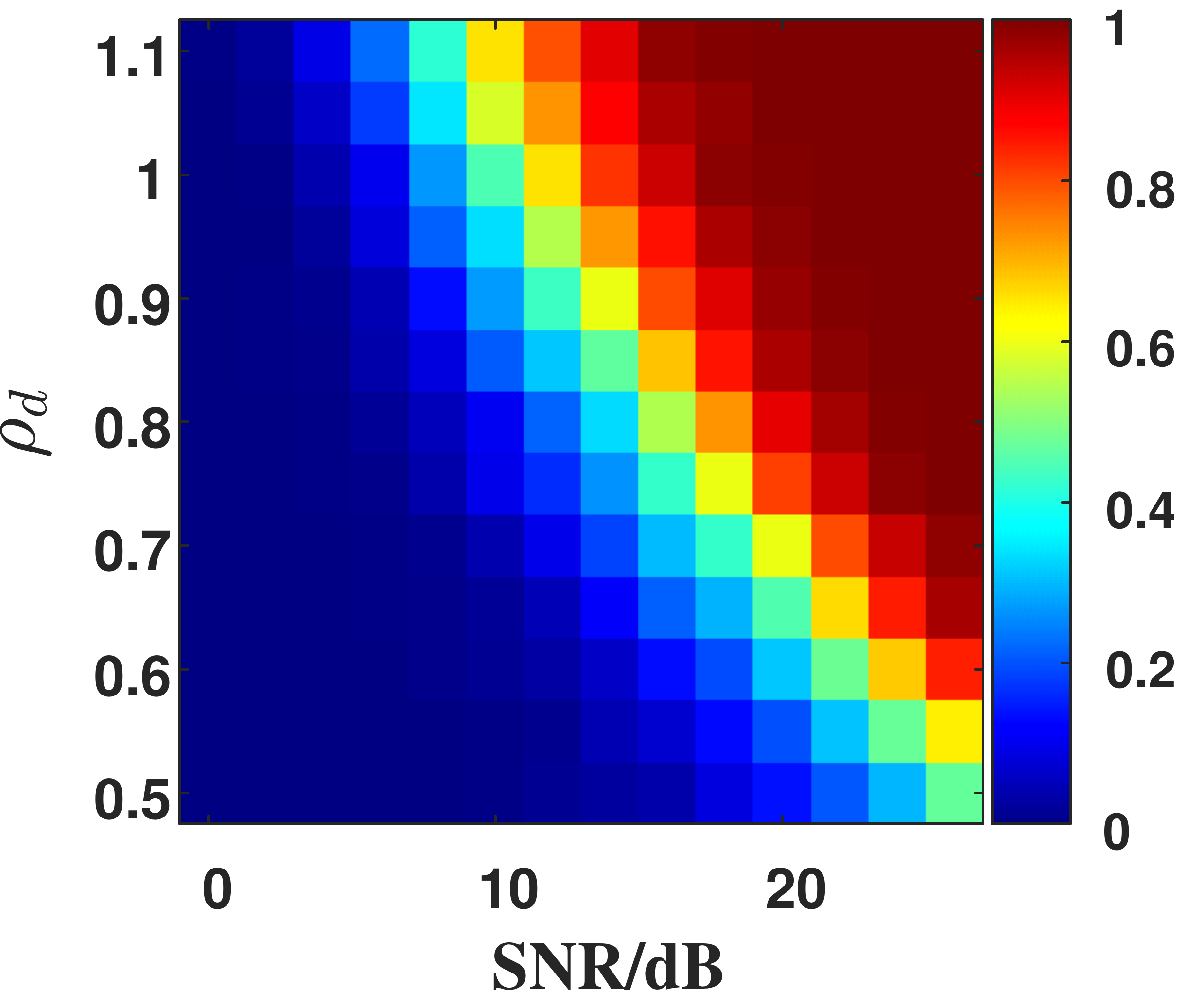}\label{MUSIC}}
  \hspace{0.05cm}
  \subfloat[]{\includegraphics[width =4.1cm, height =3.6cm]{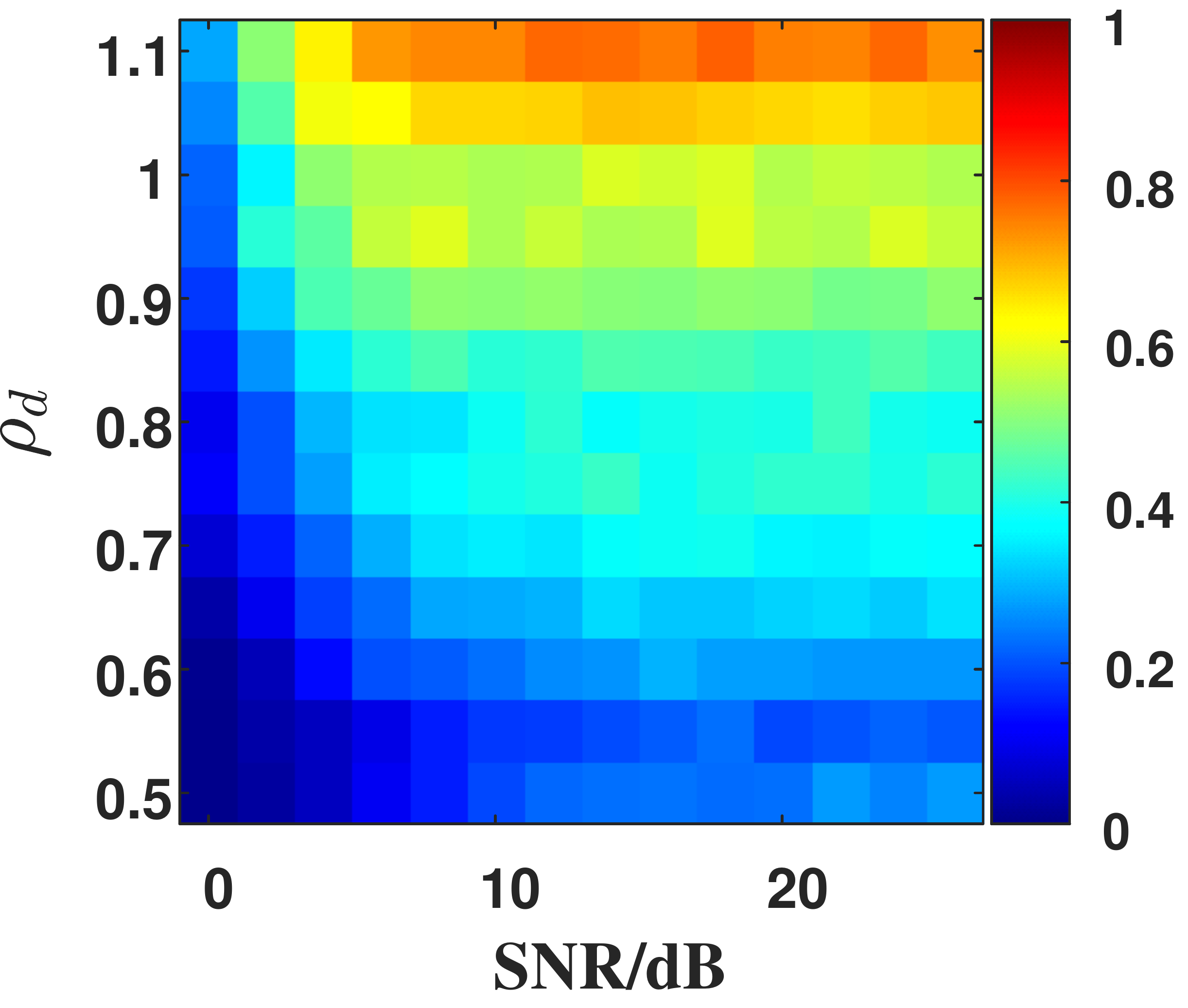}\label{OMP}}
  \hspace{0.05cm}
   \subfloat[]{\includegraphics[width =4.1cm, height =3.6cm]{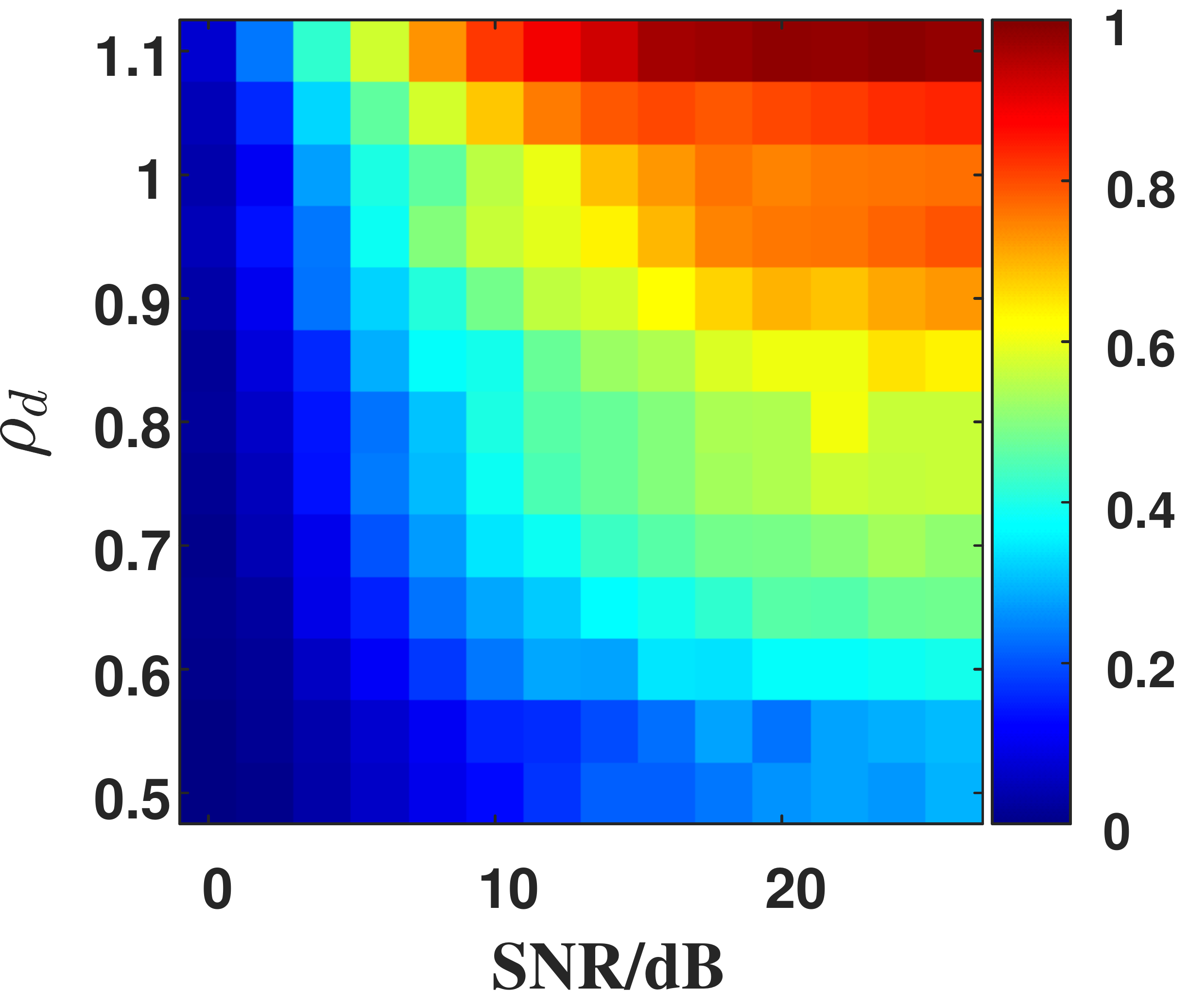}\label{ADMM}}
  \\ 
  \subfloat[]{
  \includegraphics[width =4.1cm, height =3.6cm]{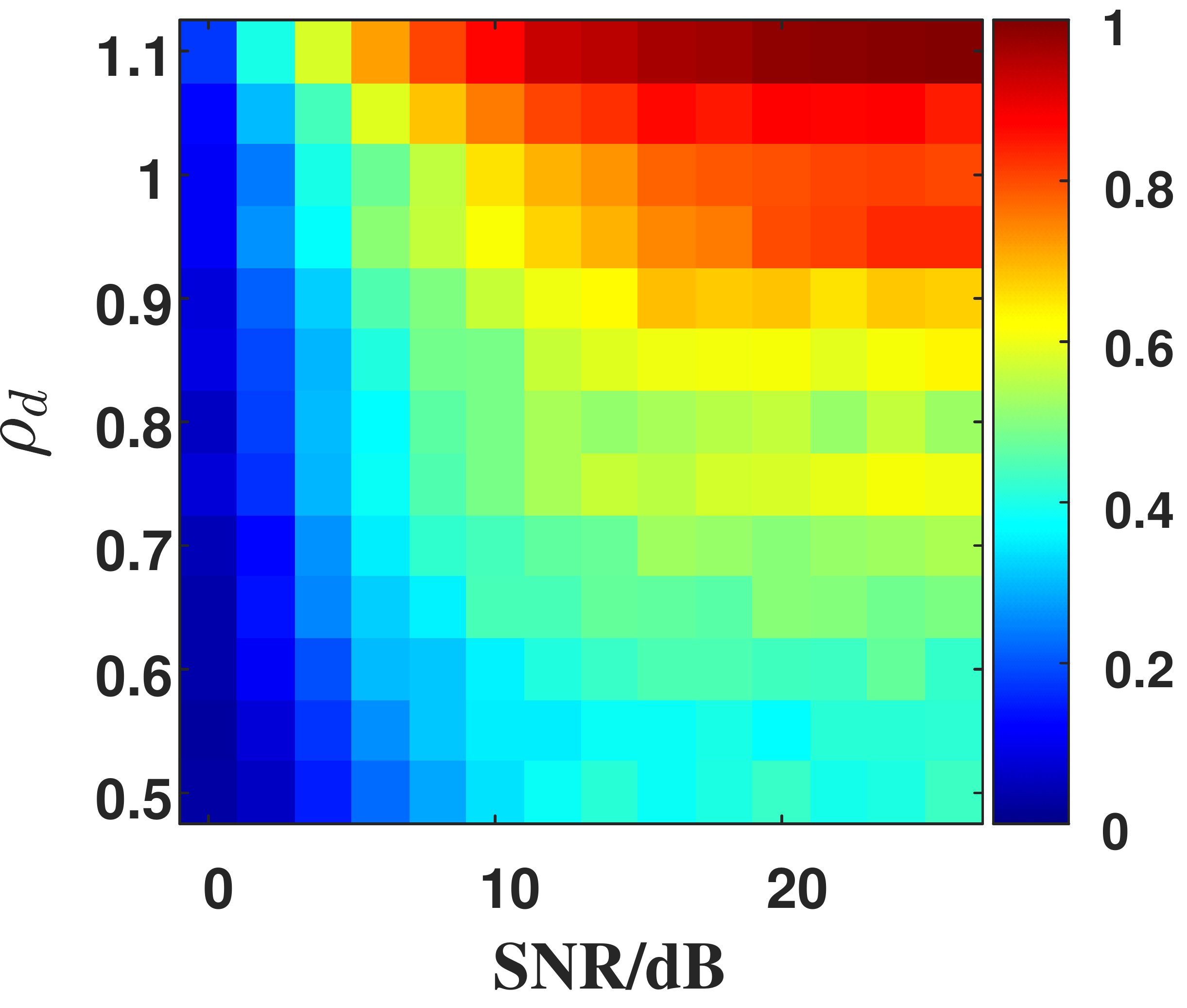}\label{ADMMnet}}
  \subfloat[]{
  \includegraphics[width =4.1cm, height =3.6cm]{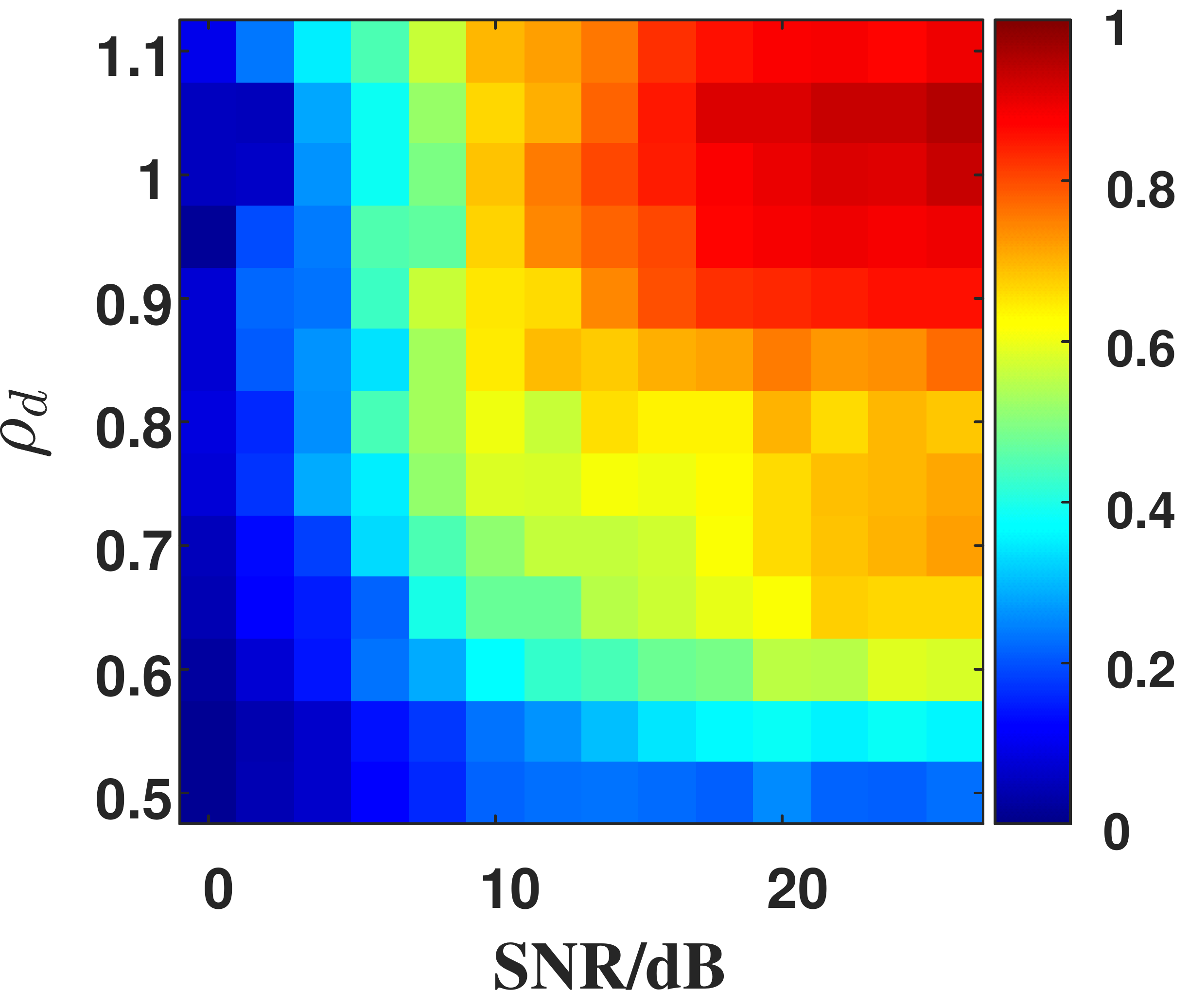}\label{DeepFreq}}
  \hspace{0.05cm}
  \subfloat[]{
  \includegraphics[width =4.1cm, height =3.6cm]{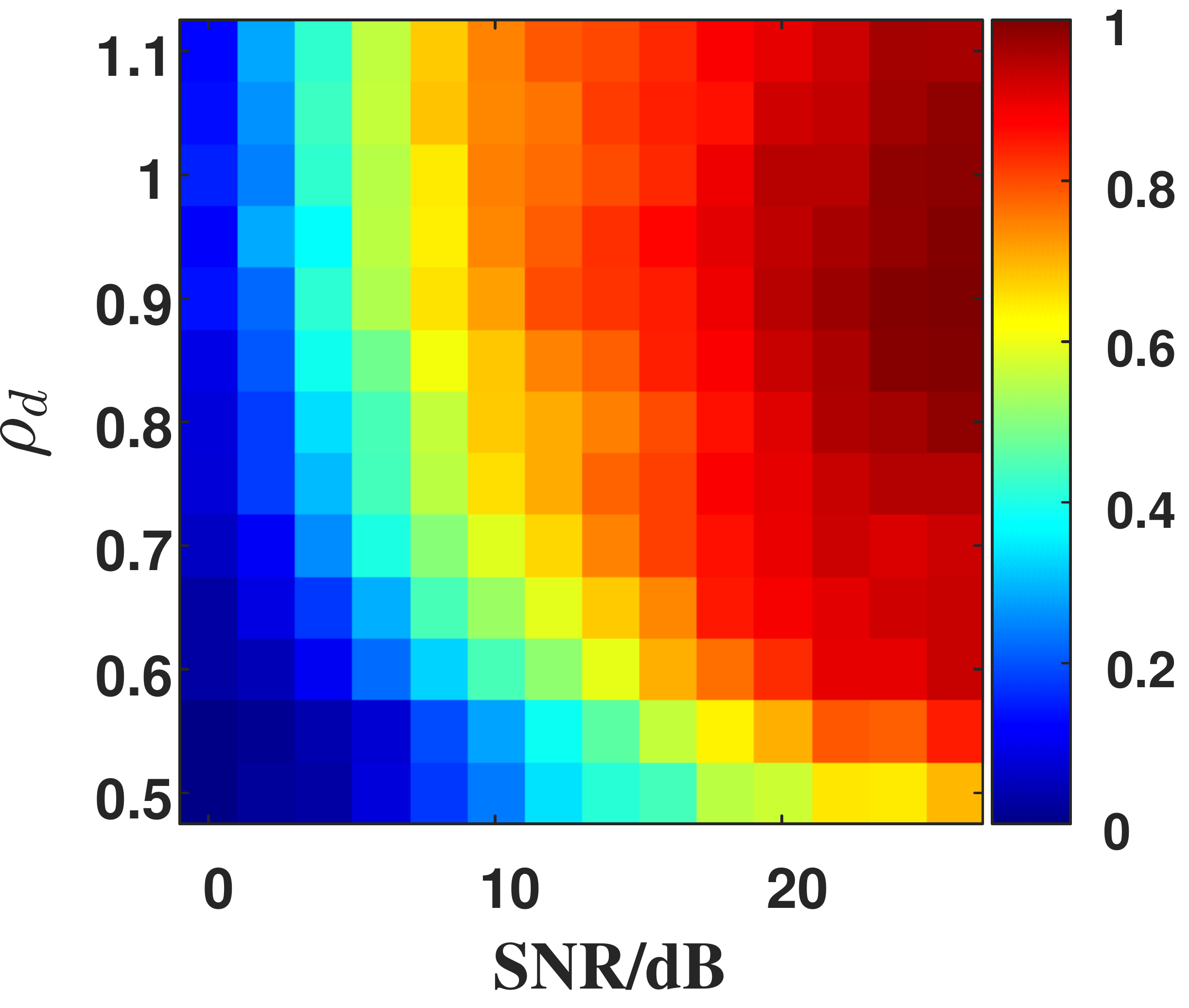}\label{CresFreq}}
  \hspace{0.05cm}
  \subfloat[]{
  \includegraphics[width =4.1cm, height =3.6cm]{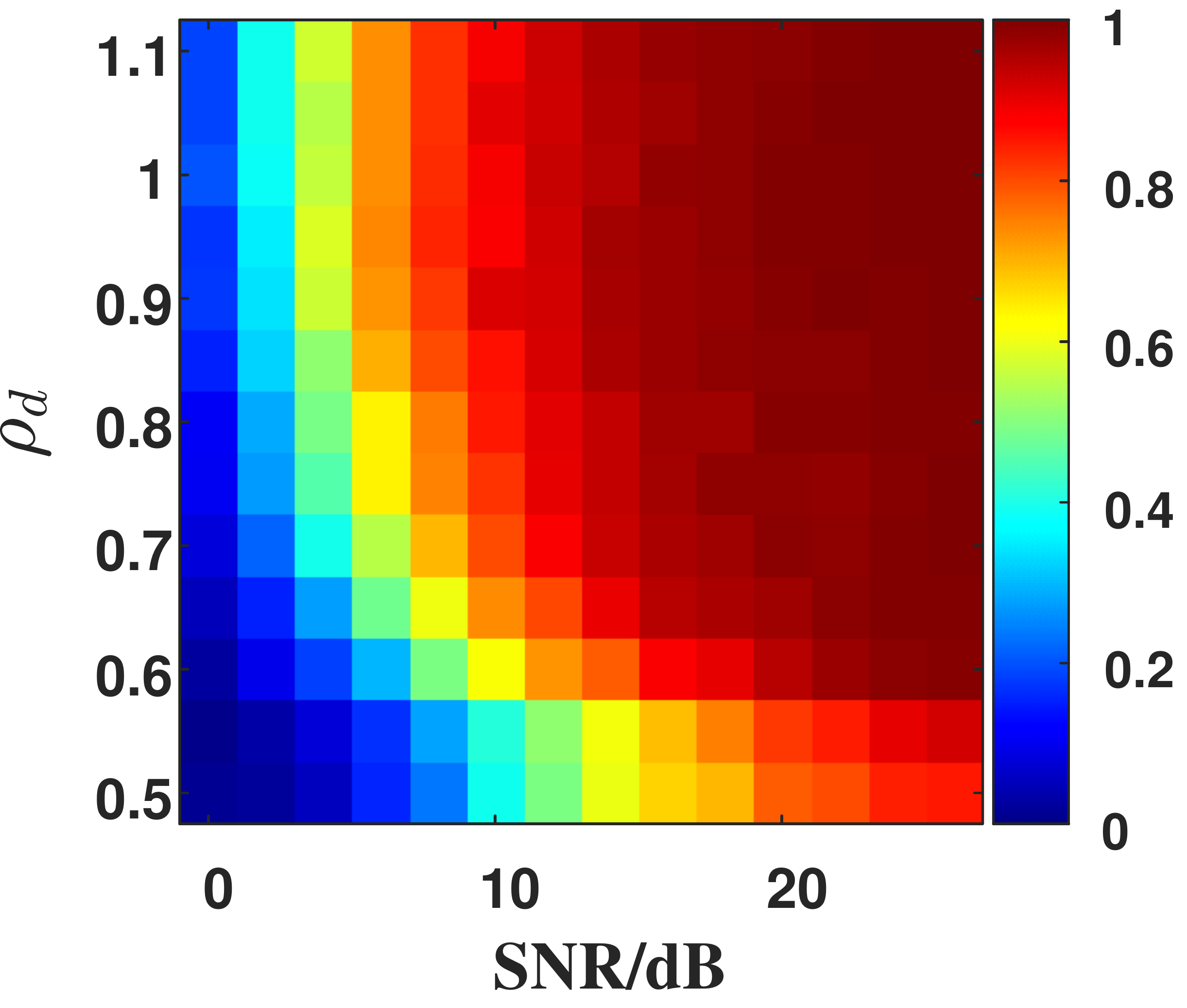}\label{DSSR-Net}}
  \vspace{0.3cm}
  \caption{Phase transition maps of a two-point super-resolution simulation experiment for different methods.
  \protect\subref{FFT}-\protect\subref{DSSR-Net} Phase transition maps based on FFT, MUSIC, OMP, ADMM, ADMM-Net, DeepFreq, CResFreq, and DSSR-Net.}
  \label{fig:phase}
\end{figure*}

Fig.~\ref{fig:phase} illustrates the results of the phase transition analysis for various methods. The color gradient that approaches red signifies a higher probability of successful super-resolution, while a gradient moving away from red indicates a lower probability. Thus, for any given method, a greater degree of red coverage reflects a superior resolution capability.

Due to the closer target, it is evident that FFT is almost ineffective. Alternatively, MUSIC relies heavily on high SNR, which allows it to better resolve closely spaced targets under such conditions. However, its resolution capability diminishes when the SNR is low.
The overall performance of OMP is superior to that of FFT, as it can successfully resolve targets even at low SNRs. However, this advantage is likely due to having prior knowledge of the number of targets. ADMM and ADMM-Net achieve a successful super-resolution at larger relative distances. However, their performance degrades for closely spaced targets, e.g., when the relative distance is less than 0.8, due to their operation along a single dimension.

Finally, both CResFreq and the proposed DSSR-Net can obtain high success resolution probability in a large range, significantly outperforming DeepFreq. However, the successful resolution performance of the proposed DSSR-Net is higher than CResFreq. 
This benefits from the proposed sparse representation model and corresponding interpretable network. It successfully achieves super-resolution RRP even at low SNRs and with closely spaced targets. DSSR-Net demonstrates a higher tolerance for both relative distance and SNR, indicating its robustness in achieving successful super-resolution due to its solid theoretical foundations that integrate both model- and data-driven methods. 
In addition, the computation time per Monte Carlo simulation of each method is shown in Table \ref{table:time}. All the methods were implemented on an NVIDIA A100 GPU with 40GB using the PyTorch framework. It can be observed that both FFT and neural network-based methods exhibit high computational efficiency, with runtimes less than 1 s. OMP shows moderate efficiency, while MUSIC and ADMM require relatively longer computation times. Notably, the neural network-based methods benefit from inherent parallelism, enabling significantly faster inference. Detailed computational complexity analysis is provided in the supplementary material.

\begin{table}[h]
  \caption{Computation Time Statistic}
    \hspace{-0.3cm}
  \begin{tabular}{ccccc}
    \toprule
    \textbf{Methods} & FFT & MUSIC & OMP & ADMM  \\
    \midrule
    \textbf{time(s)}  &  0.01&  5.10&  1.12&  22.75\\
    \midrule
    \textbf{Methods}& ADMM-Net & DeepFreq & CResFreq & DSSR-Net\\
    \midrule
    \textbf{time(s)} &0.57&  0.02&   0.03&   0.11\\
    \bottomrule
  \end{tabular}
   \label{table:time}
\end{table}

Next, we will provide a detailed analysis of the super-resolution performance under specific SNR, relative distance conditions, and in the presence of a weak target.

\begin{figure}[t]
  \subfloat[]{\includegraphics[width =8cm, height =5cm]{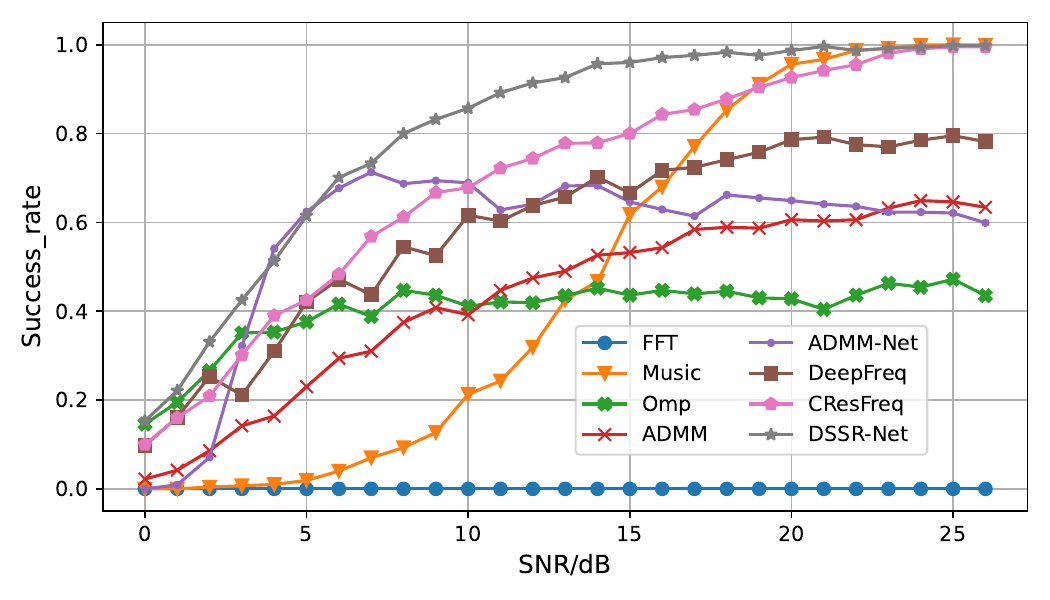}
  \label{fig:SNR_surse}}\vspace{-0.23cm}
  \\ 
  \subfloat[]{\includegraphics[width =8cm, height =5cm]{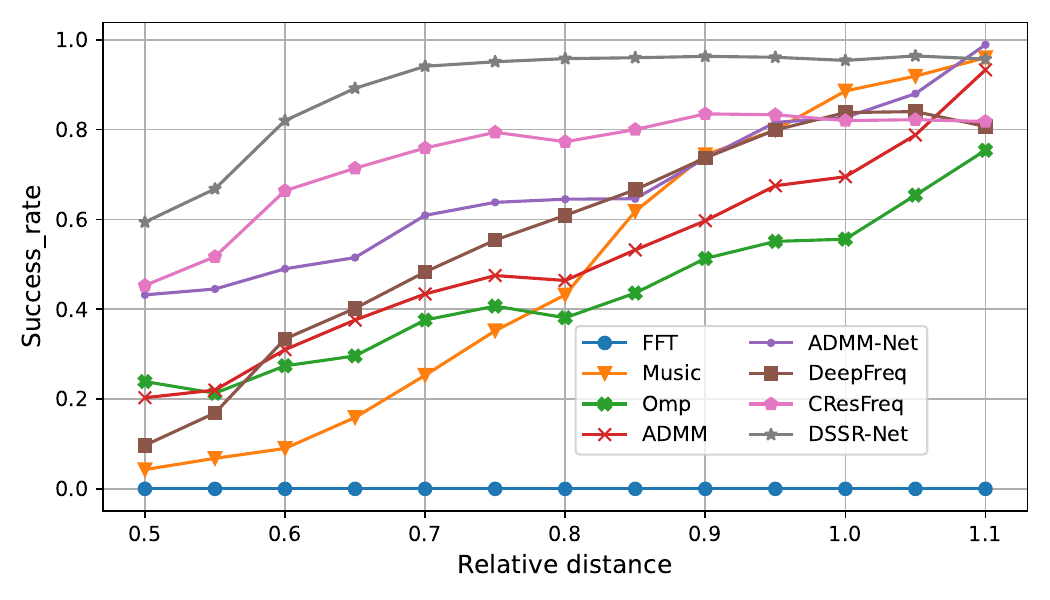}
  \label{fig:dis_surse}}\vspace{-0.23cm}
  \\
\subfloat[]{ \hspace{-0.23cm}
    \includegraphics[width =8cm, height =5cm]{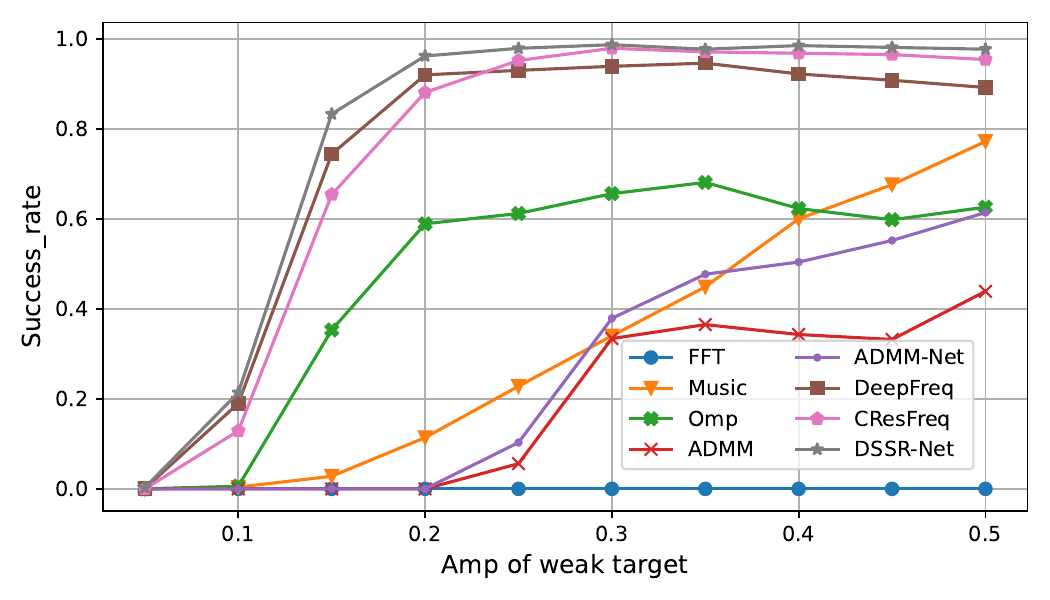}
  \label{fig:amp_surse}
  }
  \caption{The reconstruction success rate curves for different methods as SNR, relative distance, and amplitude disparity varying.
  \protect\subref{fig:SNR_surse} Success rate curves with SNR varying of ${\rho _\text{d}}=0.8$.
  \protect\subref{fig:dis_surse} Success rate curves with relative distance varying of $\mathrm{SNR}=15\,\mathrm{dB}$.
  \protect\subref{fig:amp_surse} Success rate curves with amplitude disparity varying of ${\rho _\text{d}}=1$ and $\mathrm{SNR}=20\,\mathrm{dB}$.}
  \vspace{-0.23cm}
  \label{fig:surse}
\end{figure}

\noindent \textbf{1) Performance varies with SNR, with fixed relative distance $\rho_\text{d}=0.8$}

In this subsection, a typical value of the relative distance $\rho_\text{d}=0.8$ is introduced as a baseline to analyze the impact of SNR on super-resolution performance.

\begin{figure}[t]
  \centering
  \includegraphics[width =8.5cm, height =5.5cm]{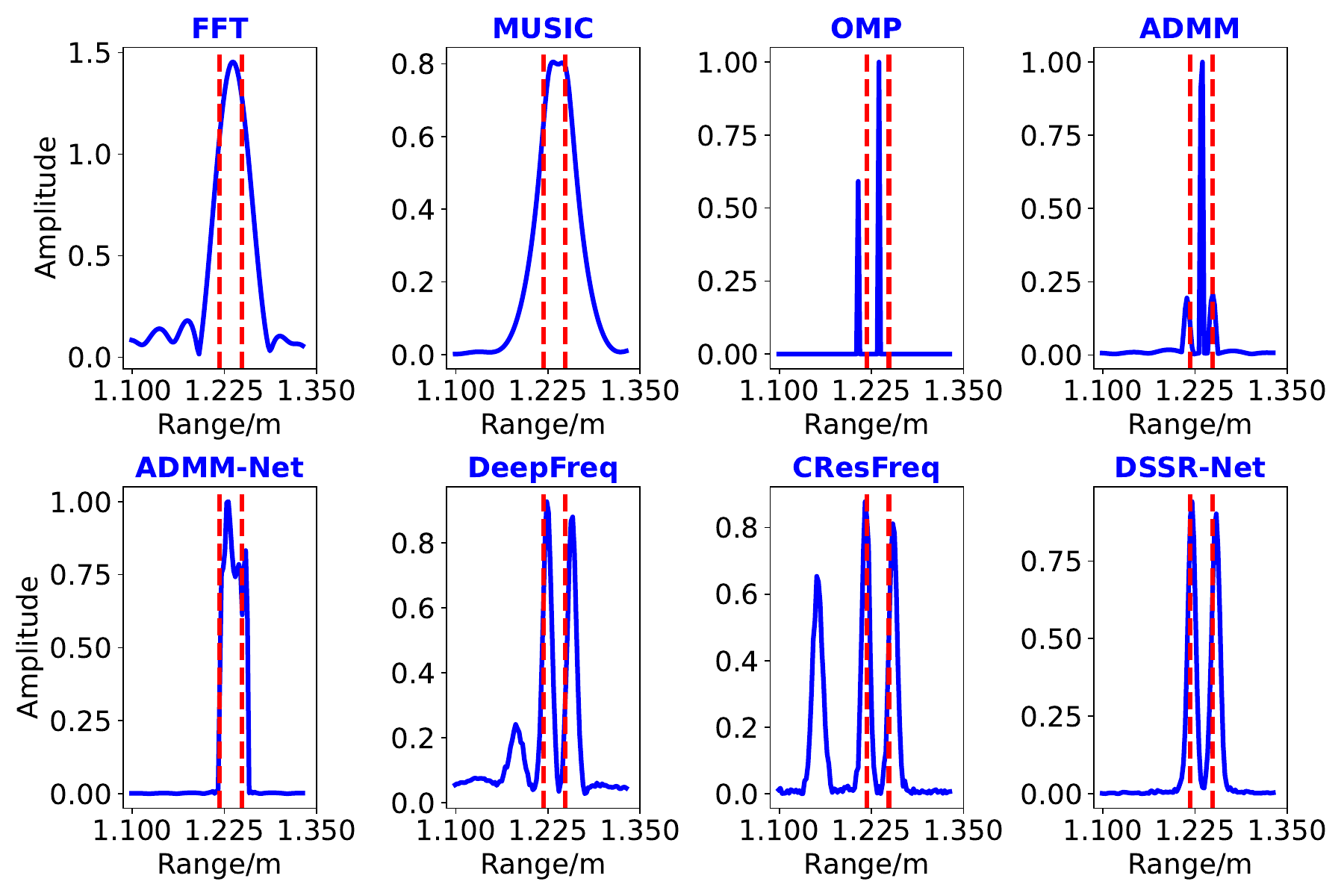}
  \vspace{-0.5cm}
  \caption{The reconstruction RRP results for different methods with SNR of $10\,\mathrm{dB}$ and ${\rho_\text{d}}=0.8$.}
  \label{fig:1d_surse}
\end{figure}

As shown in Fig.~\ref{fig:surse}\subref{fig:SNR_surse}, the success rate curves for different methods with a fixed relative distance of $\rho_\text{d}=0.8$ are presented. Furthermore, Fig.~\ref{fig:1d_surse} displays the reconstruction results of RRP with $\mathrm{SNR}=10\,\mathrm{dB}$ and $\rho_\text{d}=0.8$.

From Fig.~ \ref{fig:surse}\subref{fig:SNR_surse}, we observe that the success rates of FFT, OMP, ADMM, ADMM-Net, and DeepFreq remain relatively constant, at approximately 0$\,\%$, 44$\,\%$, 65$\,\%$, 63$\,\%$, and 80$\,\%$, respectively, as the SNR increases. The limited resolution capability of FFT is attributed to its original resolution being constrained by the bandwidth of the frequency domain.
In Fig.~\ref{fig:1d_surse}, we can see that the failure of OMP is primarily due to its absolute position offset, as the prior information only indicates the number of targets, resulting in a lower performance ceiling. Although ADMM and ADMM-Net achieve a high success rate under these conditions, they are hindered from performing at an even higher level due to a lack of detailed target priors and the constraint of operating along a single dimension. In addition,  DeepFreq is prone to generating false targets.

In contrast, the success rates of MUSIC, CResFreq, and the proposed DSSR-Net reach over 100$\,\%$ as the SNR increases. Among these methods, MUSIC has the highest SNR requirement, followed by CResFreq. The proposed DSSR-Net shows a success rate of 80$\,\%$ at $\mathrm{SNR}=8\,\mathrm{dB}$, reducing the SNR requirement by 7 to 10 $\,\mathrm{dB}$ compared to CResFreq and MUSIC. As shown in Fig.~\ref{fig:1d_surse}, although CResFreq can distinguish targets, it is also prone to interference points. The proposed DSSR-Net can alleviate this problem, which is attributed to the feature extraction module operating in a high-dimensional space, which effectively filters out noise, allowing DSSR-Net to function well under low SNR conditions. 

\noindent \textbf{2) Performance varies with relative distance ${\rho _\text{d}}$, with fixed $\mathrm{SNR}=15\,\mathrm{dB}$}

Similarly to the last subsection, a typical value of $\mathrm{SNR}=15\,\mathrm{dB}$ is used as a baseline to analyze how the relative distance ${\rho_\text{d}}$ affects the super-resolution performance in this subsection.

As shown in Fig.~\ref{fig:surse}\subref{fig:dis_surse}, the corresponding success rate curve is plotted for different methods at a fixed $\mathrm{SNR} =15\,\mathrm{dB}$. In this scenario, FFT still cannot distinguish the target, as can be seen in Fig.~\ref{fig:1d_surse}. It is stated here that FFT's failure when $\rho_\text{d}>1$ is due to noise and the strict definition of successful super-resolution.

The remaining seven methods demonstrate relatively high success rates for super-resolution when the targets are further apart. However, since this paper emphasizes super-resolution performance, it focuses mainly on situations where the targets are closer than the resolution limit. In a range below a resolution unit, the success super-resolution rate of the proposed DSSR-Net is higher than that of other methods. For example, at a relative distance ${\rho _\text{d}}=0.7$, the proposed DSSR-Net has a success rate 65$\,\%$, 52$\,\%$, 42$\,\%$, 32$\,\%$, 38$\,\%$, and 10$\,\%$ higher than MUSIC, OMP, ADMM, ADMM-Net, DeepFreq, and CResFreq, respectively. This is because, compared to low-dimensional space processing, the high-dimensional space in DSSR-Net helps distinguish the differential features of closely spaced targets, thereby achieving successful super-resolution.

\noindent \textbf{3) Performance varies with amplitude disparity between targets, with fixed relative distance $\rho_\text{d} = 1$ and $\mathrm{SNR}=20\,\mathrm{dB}$}

This subsection focuses on assessing the algorithm’s ability to achieve super-resolution in the presence of weak targets. In this experiment, two targets are placed at a relative distance $\rho_\text{d}=1$ and $\mathrm{SNR}=20\,\mathrm{dB}$. One target has a constant amplitude of 1, while the other is a weak target with its amplitude varying from 0.05 to 0.5. The evaluation was conducted using the same definition of successful super-resolution as described previously.

As shown in Fig.~\ref{fig:surse}\subref{fig:amp_surse}, the corresponding success rate curves are plotted for different methods. It can be observed that, compared with other methods, DeepFreq, CResFreq, and the proposed DSSR-Net generally achieve better super-resolution performance for weak targets. Notably, when the weak target amplitude is 0.15, the proposed method achieves a success rate exceeding 80$\,\%$, which is 8$\,\%$ and 18$\,\%$ higher than that of DeepFreq and CResFreq, respectively, and over 30$\,\%$ better than those of the remaining methods. These results demonstrate the superiority of the proposed method for super-resolution estimation of frequency in scenarios involving one strong and one weak closely-spaced targets. It can be attributed to its use of a high-dimensional space and an analytically designed filtering network.

\begin{figure}[t]
  \centering
  \vspace{-0.2cm}
  \includegraphics[width =8.5cm, height =5.8cm]{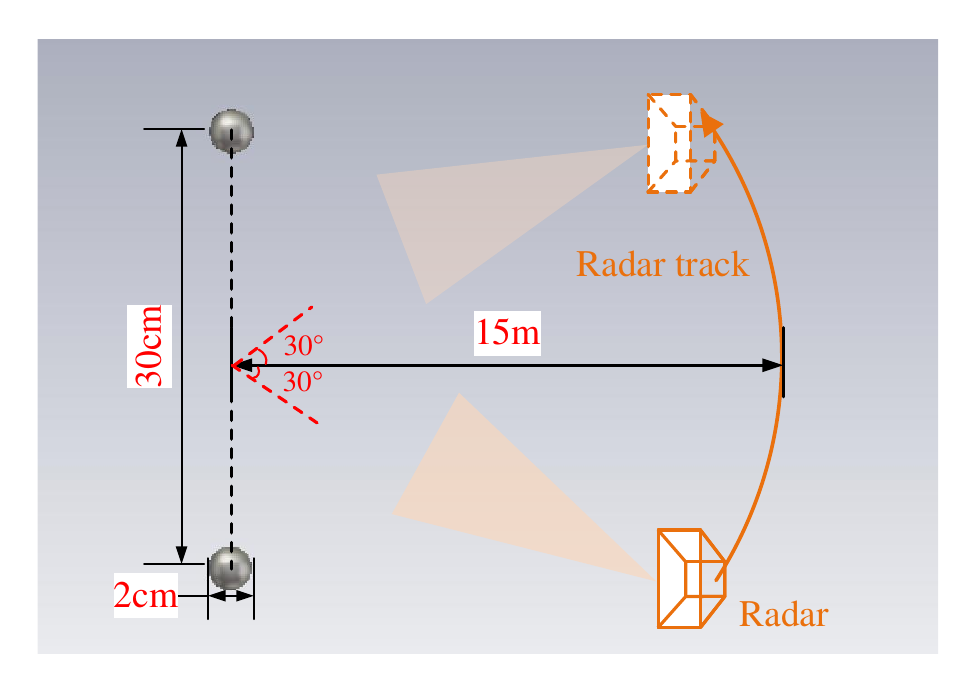}
  \caption{Radar observation geometry of position movement and targets' position in the CST experiment.}
  \label{fig:CSTjihe}
\end{figure}

\begin{figure*}[ht]
  \vspace{0.4cm}
  \subfloat[]{\includegraphics[width =17.6cm, height =2.6cm]{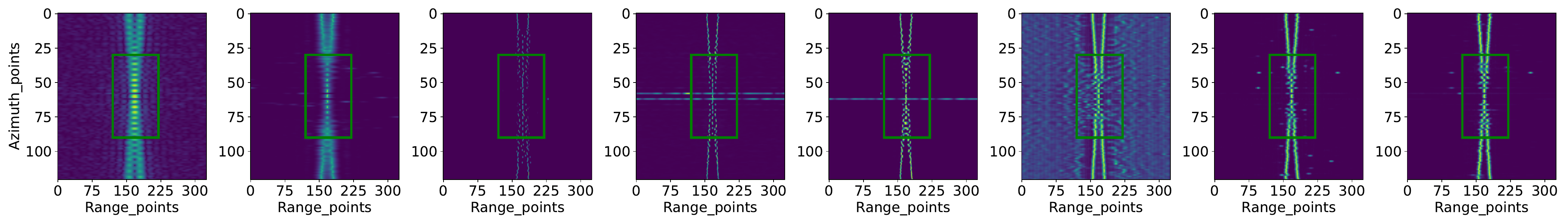}
  \label{fig:range_cst_change1}}
  \\
  \subfloat[]{\hspace{0.1cm}\includegraphics[width =17.6cm, height =2.6cm]{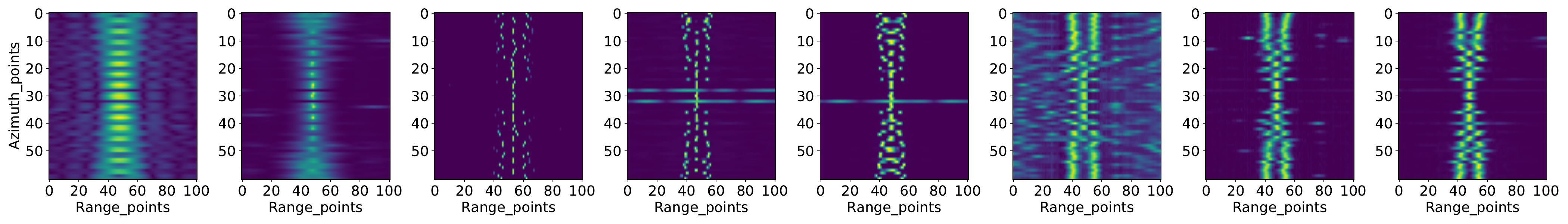}
  \label{fig:range_cst_change2}}\vspace{-0.2cm}
  \\
  \subfloat[]{ \hspace{-0.4cm}\vspace{-0.2cm}
  \includegraphics[width =17.8cm, height =2.8cm]{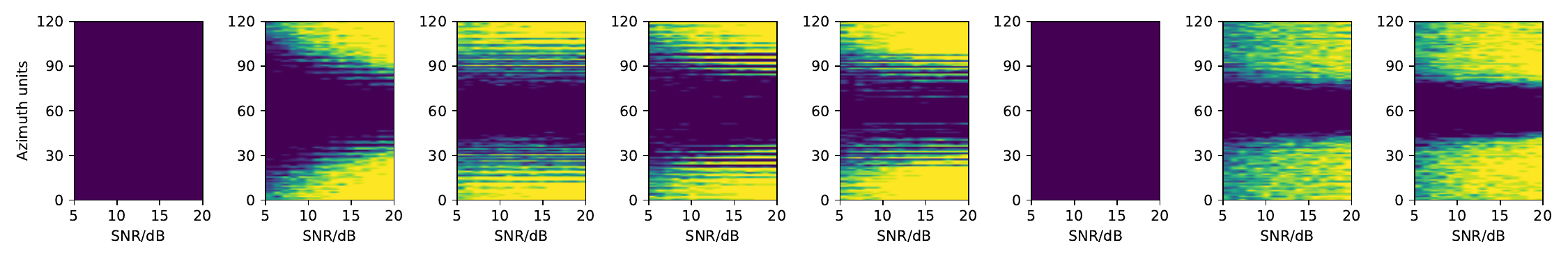}
  \label{fig:range_cst_change3}}
  \caption{One-dimensional reconstruction results of the different methods and orientation views in the CST experiment. From left to right correspond respectively to FFT, MUSIC, OMP, ADMM, ADMM-Net, DeepFreq, CResFreq, and the proposed DSSR-Net.
  \protect\subref{fig:range_cst_change1} One-dimensional reconstruction results of different methods at $\mathrm{SNR}=15\,\mathrm{dB}$.
  \protect\subref{fig:range_cst_change2} The enlarged images of the green boxes in \protect\subref{fig:range_cst_change1} of the different methods.
  \protect\subref{fig:range_cst_change3} The successful super-resolution maps and the variation of SNR.}
  \label{fig:range_cst_change}
\end{figure*}

\begin{figure}[t]
  \centering
  \vspace{-0.6cm}
  \includegraphics[width =8cm, height =5.8cm]{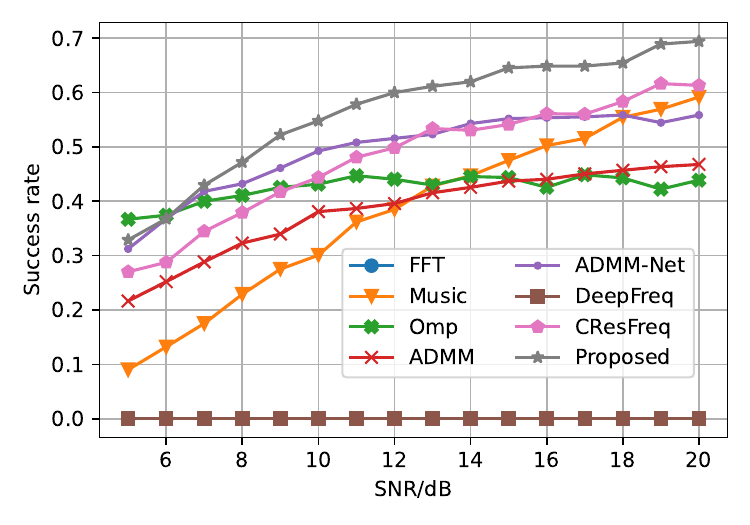}
  \vspace{-0.3cm}
  \caption{The statistical curve of successful discrimination probability with the variation of SNR in the CST experiment.}
  \label{fig:acc_table_cst}
\end{figure}
\subsection{Experiments with electromagnetic simulation data}

In this section, the effectiveness of the proposed DSSR-Net is validated using electromagnetic simulation data from two spheres in the CST MICROWAVE STUDIO. The observation geometry is illustrated in Fig.~\ref{fig:CSTjihe}, using the circular trajectory SAR mode, with rotation angles of up to 30° in both up and down directions. The radar bandwidth and carrier frequency are $1.92\,\mathrm{GHz}$ and $14\,\mathrm{GHz}$, respectively.

Figs.~\ref{fig:range_cst_change}\subref{fig:range_cst_change1} and \ref{fig:range_cst_change}\subref{fig:range_cst_change2} show the results of super-resolution RRPs and the enlarged images of the green boxes of the different methods in different orientation views at an SNR of $ 15\,\mathrm{dB}$. It can be seen that FFT and MUSIC often fail to distinguish distance curves in most cases due to bandwidth limitations.
OMP, ADMM, and ADMM-Net demonstrate superior performance compared to MUSIC, yet their results remain unsatisfactory. DeepFreq demonstrates fundamental detection capability for true targets but exhibits substantial false alarm artifacts. 
In addition, compared with CResFreq, the proposed DSSR-Net has significantly fewer interference points, which is due to the proposed interpretable feature extraction module.

As shown in Fig.~\ref{fig:range_cst_change}\subref{fig:range_cst_change3} for the successful super-resolution probability and the variation of SNR. The definition of a successful super-resolution is based on the criteria described in Sec~\ref{section4.1}, where the yellow color represents a successful super-resolution and another color represents a failed super-resolution. It can be observed that FFT, limited by bandwidth limitations, fails to discriminate when the intermediate targets are closer. MUSIC can effectively distinguish closely spaced targets in high SNR conditions, but its probability of achieving successful super-resolution decreases in low SNR scenarios. OMP, ADMM, and ADMM-Net demonstrate relatively consistent performance across all SNR levels, though they struggle to discriminate targets that are closer together. DeepFreq produces unsatisfactory results due to weak spurious targets that lead to detection failures. Although the gap between CResFreq and the proposed DSSR-Net in the successfully distinguished area is small, the proposed DSSR-Net shows a higher success super-resolution probability.

Fig.~\ref{fig:acc_table_cst} illustrates the statistical curve of the successful super-resolution probability with the variation of the SNR. It can be observed that the proposed DSSR-Net reaches a 60$\,\%$ successful super-resolution probability at about $ 12\,\mathrm{dB}$, while CResFreq and MUSIC require about $ 19\,\mathrm{dB}$ and $ 20\,\mathrm{dB}$, respectively. The other methods do not reach this level. Similar to the phase transition analysis in Sec~\ref{section4}-\ref{section4.1}, the proposed DSSR-Net demonstrates a higher success probability even under low SNR conditions.
At SNR reaches $20\,\mathrm{dB}$, both the proposed DSSR-Net, CResFreq, and MUSIC struggle to discriminate effectively due to the close proximity of the targets.

\subsection{Experiments with the real data}

\begin{figure*}[htbp]
  \centering
  \vspace{-0.2cm}
  \subfloat[]{\includegraphics[width =8.8cm, height =4.5cm]{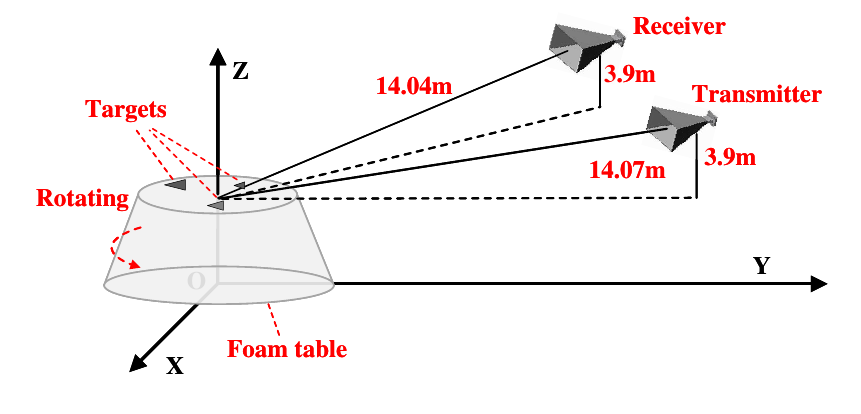}
  \label{fig:anshijihe}}
  \subfloat[]{\includegraphics[width =8.8cm, height =5.1cm]{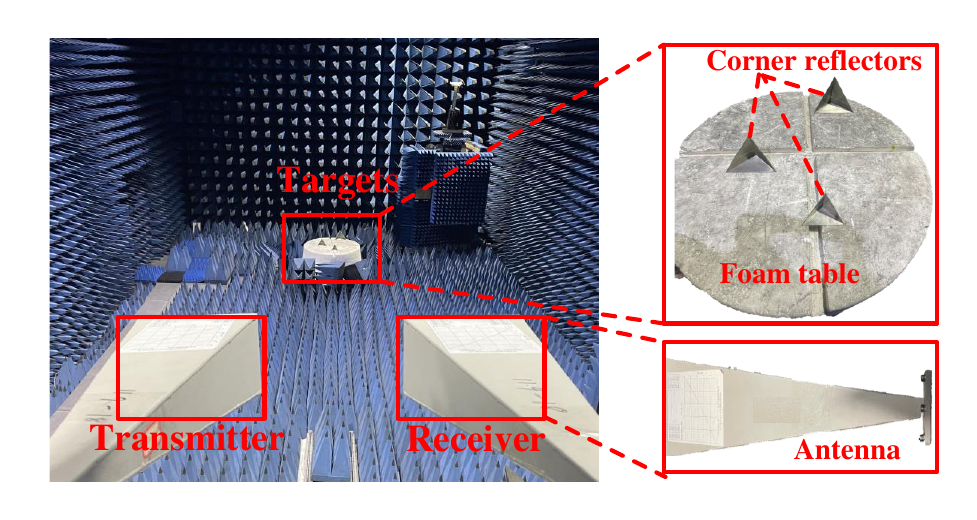}
  \label{fig:anshishiwu}}

  \caption{The radar observation geometry and real-world images of the scenarios in an anechoic chamber experiment.
  \protect\subref{fig:anshijihe} Radar observation geometry of the anechoic chamber experiment.
  \protect\subref{fig:anshishiwu} Real-world images of the scenarios of the anechoic chamber experiment.
  }
  \label{fig:anshi}
\end{figure*}

\begin{figure*}[htbp]
  \centering
  \subfloat[]{\includegraphics[width =17.5cm, height =2.5cm]{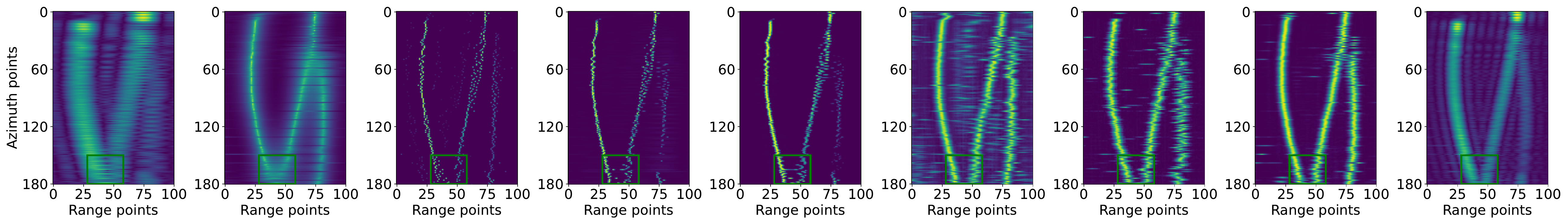}
  \label{fig:range_anshi_change1}}
  \\
  \subfloat[]{\includegraphics[width =17.5cm, height =2.5cm]{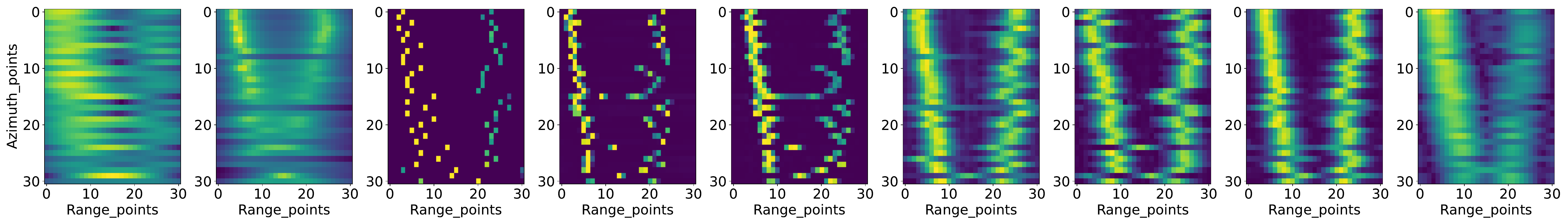}
  \label{fig:range_anshi_change2}}
  \caption{The one-dimensional reconstruction results of the different methods and orientation views in simulation experiment. From left to right, the experimental outcomes correspond respectively to FFT, MUSIC, OMP, ADMM, ADMM-Net, DeepFreq, CResFreq, the proposed DSSR-Net and FFT with double the bandwidth.
  \protect\subref{fig:range_anshi_change1} The global reconstruction results of different methods.
  \protect\subref{fig:range_anshi_change2} The local images of the green boxes in \protect\subref{fig:range_anshi_change1} of the different methods.}
  \label{fig:range_anshi}
\end{figure*}

In this section, we present a real-world radar data collected from three corner reflectors within a microwave anechoic chamber to verify the effectiveness of the proposed DSSR-Net. The radar observation setup and the corresponding images of the scenarios are illustrated in Figs.~\ref{fig:anshi}\subref{fig:anshijihe} and \ref{fig:anshi}\subref{fig:anshishiwu}, respectively. 

The experiment in the anechoic chamber uses a bistatic configuration, featuring one radar antenna as a transmitter and another as a receiver. Both antennas are mounted at a height and directed downward. The target area consists of three corner reflectors arranged on a foam turntable. By controlling the rotation of the turntable, we can modify the slant range of the targets. The key parameters of the observation geometry and radar system are summarized in Table~\ref{table:anshi}. Furthermore, to validate the effectiveness of super-resolution, we also collected data with double the bandwidth, which was subsequently processed using FFT.

\begin{table}[h]
  \caption{The primary experiment parameters in microwave anechoic chamber}   \centering
  \begin{tabular}{cc}
  \toprule
    \textbf{Parameter} & \textbf{Value} 	 \\
    \midrule
    Carrier frequency &   $14.3\,\mathrm{GHz}$\\
    Bandwidth &   $640\,\mathrm{MHz}$\\
    Frequency points &  64   \\
    Elevation angle& 16.3°\\
    Azimuth angle& [-15°, 15°]\\
  \bottomrule
  \end{tabular}
  \label{table:anshi}
\end{table}

The experimental data results for handling various orientations in the anechoic chamber are shown in Fig.~\ref{fig:range_anshi}. Here, Fig.~\ref{fig:range_anshi}\subref{fig:range_anshi_change1} represents the global image, while Fig.~\ref{fig:range_anshi}\subref{fig:range_anshi_change2} denotes the local image within the green box in Fig.~\ref{fig:range_anshi}\subref{fig:range_anshi_change1}, used to highlight differences in super resolution performance.

As shown in Fig.~\ref{fig:range_anshi}\subref{fig:range_anshi_change1}, it can be observed from all methods that the distances between the three corner reflectors vary with the change in azimuth angle. However, because of differences in resolution capabilities, the local targets at closer ranges cannot be distinguished. FFT can only separate targets that are farther apart. MUSIC also fails to distinguish closer targets. OMP can separate closer targets, but target estimation is inaccurate and prone to random shifts. ADMM and ADMM-Net, similar to OMP, are prone to random shifts at closer ranges and easily filter out weak targets. DeepFreq can separate targets, but has higher sidelobes, leading to false targets. 
In most cases, the proposed DSSR-Net successfully distinguishes targets with minimal displacement error while generating significantly fewer spurious targets than CResFreq, demonstrating superior performance among the various methods.

Furthermore, from the images in Fig.~\ref{fig:range_anshi}\subref{fig:range_anshi_change2}, it can be observed that FFT and MUSIC fail to distinguish the targets in this area. OMP, ADMM, and ADMM-Net can outline the distance variations to some extent, but they exhibit significant oscillations that do not align with the essence of the slant range gradient. 
DeepFreq and CResFreq can separate targets, but some points experience displacement, and the slant range system curve is not accurately represented. In comparison to FFT results with double the bandwidth, where the slant range variation line on the right should exhibit partial slope,  DeepFreq and CResFreq exhibit higher variability in their estimations, whereas the proposed DSSR-Net demonstrates superior stability and reliably captures the gradual change characteristics, yielding more accurate and consistent distance measurements.

Through this anechoic chamber experiment, it is demonstrated that the proposed DSSR-Net can be applied to real data for super-resolution RRP. It achieves high-quality RRP results using a model-guided and multi-dimensional data-driven network framework and can aid subsequent applications.
\section{Conclusion}
\label{section5}
In this paper, a Super-Resolution Network with Dimension Scaling (DSSR-Net) is proposed to obtain the super-resolution RRP. 
The entire network is designed with the guidance of a sparse representation model which lifts the echo into a high-dimensional space;
thus, its interpretability is improved. Numerical simulations and real-data experiments demonstrate that the proposed DSSR-Net obtains better resolution and performs well at relatively lower SNR scenarios compared to other state-of-the-art methods. 
In this work, one-dimensional (1D) super-resolution reconstruction problem is addressed. In principle, the same idea used in DSSR-Net could be extended to 2D and 3D scenarios, which will be conducted in the future. 

\ifCLASSOPTIONcaptionsoff
  \newpage
\fi

\bibliography{ref}
\bibliographystyle{IEEEtran}

\begin{IEEEbiography}[{\includegraphics[width=1in,height=1.25in,clip,keepaspectratio]{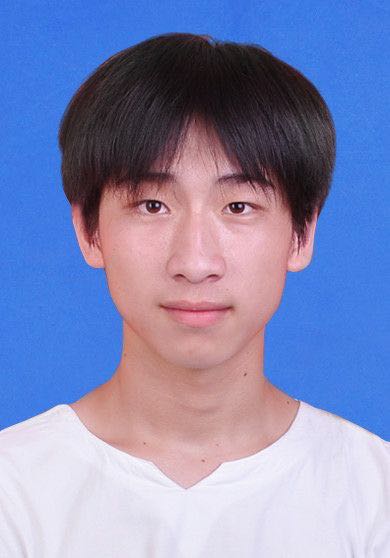}}]
{Ziwen Wang}{\space} was born in Jiangxi, China, in 1999. He received the B.Sc. degree in communication engineering from the Northwestern Polytechnical University, Xi’an, China, in 2020, and the M.Sc. degree in Information and Communication Engineering from the Beijing Institute of Technology, Beijing, in 2023. He is currently pursuing the Ph.D. degree in Information and Communication Engineering with the Beijing Institute of Technology, Beijing.

His research interests mainly include radar signal processing, sparse aperture ISAR imaging, and deep learning.
\end{IEEEbiography}%

\begin{IEEEbiography}
[{\includegraphics[width=1in,height=1.25in,clip,keepaspectratio]{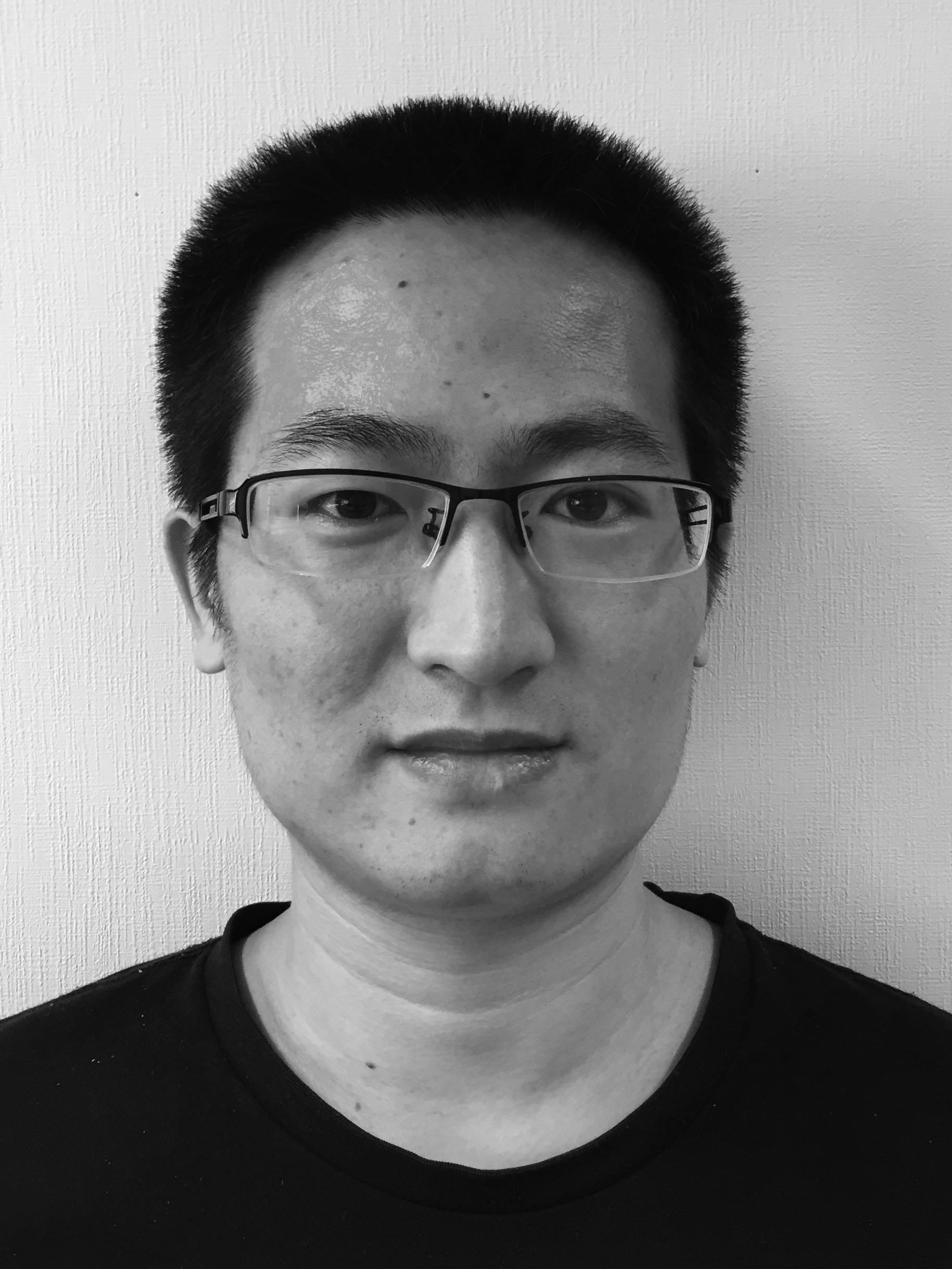}}]
{Jianping Wang} (Member, IEEE) received the Ph.D. degree in electrical engineering from the Delft University of Technology, Delft, The Netherlands, in 2018. From August 2012 to April 2013, he was a Research Associate at the University of New South Wales, Sydney, NSW, Australia, with a focus on frequency-modulated continuous wave synthetic aperture radar signal processing for formation flying satellites. He was a Post-Doctoral Researcher and a Guest Researcher with the Group of Microwave Sensing, Signals and Systems (MS3), Delft University of Technology, from 2018 to 2024. Since 2024, he has been with the School of Information and Electronics, Beijing Institute of Technology, Beijing, China. 

His research interests include microwave imaging, signal processing, and antenna array design. Dr. Wang was a TPC Member of the IET International Radar Conference, Nanjing, China, in 2018 and 2023. He was a Finalist for the Best Student Paper Award in the International Workshop on Advanced Ground Penetrating Radar (IWAGPR), Edinburgh, U.K., in 2017, and the International Conference on Radar, Brisbane, Australia, in 2018. He has served as a reviewer of many IEEE journals.
\end{IEEEbiography}

\begin{IEEEbiography}
[{\includegraphics[width=1in,height=1.25in,clip,keepaspectratio]{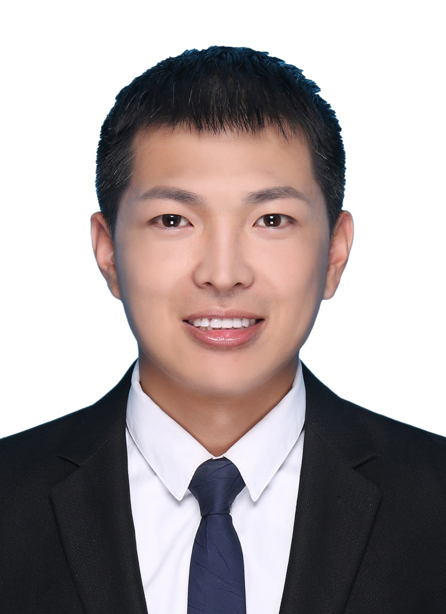}}]
{Pucheng Li} (Member, IEEE) received the B.Eng. and M.Eng. degrees in electronic engineering from the Civil Aviation University of China, Tianjin, China, in 2017 and 2020, respectively, and the Ph.D. degree in Electronic and Information Engineering from the Beijing Institute of Technology (BIT), Beijing, China, in 2024. 

He is currently a Postdoctoral Fellow with the Radar Technology Research Institute at BIT. His research focuses on high-resolution radar imaging and machine learning.
\end{IEEEbiography}

\begin{IEEEbiography}
[{\includegraphics[width=1in,height=1.25in,clip,keepaspectratio]{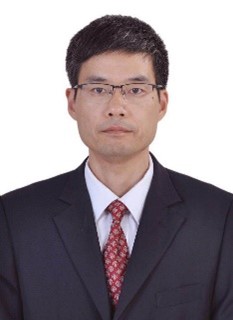}}]
{Zegang Ding} (Senior Member, IEEE) received the Ph.D. degree from the Beijing Institute of Technology (BIT), Beijing, China, in 2008. In 2008, he was a Visiting Scholar with the Communication Group, School of Electronic and Electrical Engineering, University of Birmingham, Birmingham, U.K. He is currently a part of the Teaching Staff at the Department of Electronic Engineering, BIT. 

His research interests include synthetic aperture radar imaging and system design. 
\end{IEEEbiography}

\end{document}